\documentclass[preprint,5p,times,twocolumn]{elsarticle}
\usepackage{times,url,comment}
\usepackage{amsmath,amssymb,amsfonts,bbm}
\usepackage[tight,footnotesize]{subfigure} 
\usepackage{epsfig,epsf}
\usepackage{array}
\usepackage{longtable}
\usepackage{multirow}
\usepackage{graphicx} 
\usepackage{mdwmath}
\usepackage{mdwtab}
\usepackage{eqparbox}
\usepackage{booktabs}
\usepackage{ctable}
\usepackage{eso-pic}
\usepackage[font=small,labelfont=bf]{caption}
\usepackage{breqn}
\usepackage{float}
\usepackage{longtable}
\usepackage{enumerate}

\definecolor{DG}{rgb}{0,0.4,0}
\newcommand*\rot{\rotatebox{90}}

\begin{document}
\begin{frontmatter}

%%%%%%%%%%%%%%%%%%%%%%%%%%%%%%%%%%%%%%%%%%%%%%%%%%%%%%%%%%%%%%

\title{Self-Calibration Methods for Uncontrolled Environments in Sensor Networks: A Reference Survey}

\author[upc]{Jose M. ~Barcelo-Ordinas\corref{upccor1}\fnref{upcL1}}
\ead{joseb@ac.upc.edu}
\author[utc]{Messaud ~Doudou}
\ead{messaoud.doudou@hds.utc.fr}
\author[upc]{Jorge ~Garcia-Vidal\fnref{upcL1}}
\ead{jorge@ac.upc.edu}
\author[cerist]{Nadjib ~Badache}
\ead{badache@cerist.dz}
\cortext[upccor1]{Corresponding author}
\fntext[upcL1]{Declarations of interest: none}
%\fntext[upcL1]{This work is supported by the National Spanish funding TIN2016-78473-C3-1-R, regional project 2017-SGR990, and 
%  H2020 CAPTOR project.}
\address[upc]{Universitat Politecnica de Catalunya, 08034 Barcelona, Spain}
\address[utc]{Sorbonne universités, Université de technologie de Compiègne, Heudiasyc Lab., Compiègne, France}
\address[cerist]{CERIST Research Center, Algiers, Algeria}

%\version{1.20}
%\date{\today}
%\maketitle

\begin{abstract}
Growing progress in sensor technology has constantly expanded the number and range of 
low-cost, small, and portable sensors on the market, increasing the number and type of
physical phenomena that can be measured with wirelessly connected sensors. 
Large-scale deployments of wireless sensor networks (WSN) involving 
hundreds or thousands of devices and limited budgets often constrain the choice 
of sensing hardware, which generally has reduced accuracy, 
precision, and reliability. Therefore, it is challenging to achieve good data
quality and maintain error-free measurements during the whole system lifetime.
Self-calibration or recalibration in ad hoc sensor networks to preserve data quality 
is essential, yet challenging, for several reasons, such as the existence of random 
noise and the absence of suitable general models. Calibration performed in the field,
without accurate and controlled instrumentation, is said to be in an uncontrolled environment.
This paper provides current and fundamental self-calibration approaches and models
for wireless sensor networks in uncontrolled environments.
\end{abstract}


\end{frontmatter}
%%%%%%%%%%%%%%%%%%%%%%%%%%%%%%%%%%%%%%%%%%%%%%%%%%%%%%%%%%%%%%

\section{Introduction}
\label{Sec:1}
% Context
In recent years there has been a growing interest in the use of networks of monitoring
nodes based on low-cost sensor technology. The monitored parameters measured by sensors include
air quality parameters (e.g., concentration of gases such as NO$_2$, NO, CO, CO$_2$, and O$_3$, or concentration
of particulate matter, such as PM2 or PM10), environmental parameters (e.g., temperature, humidity, radioactivity),
noise, wind, tracking of objects or persons, presence of animals, and perimetric security. 
Many of these sensors are sold by manufacturers without individual calibration parameters except
for some generic calibration values in the data-sheet description of sensor devices.
In some situations, the sensing devices were individually calibrated by the manufacturer, usually under
controlled conditions. In any case, calibration, either during the pre-installation phase or
during the monitoring network operation, is necessary in order to achieve the data quality
requirements.

% why it is important
Sensor calibration in the deployment of large-scale cyber-physical IoT
systems and ad hoc sensor networks is mandatory due to the inherent process of 
device imperfection and noise in the massive data collected. In recent years 
there has been a growing interest in applying all the theoretical knowledge 
acquired in wireless sensor networks (WSN) to commercial or research 
deployments. Much of this research has focused on challenges related to 
communication protocols suited to WSN and the management 
of their energy consumption \cite{pantazis2007,huang2013,pantazis2013,barcelo2013survey}. From the 
very beginning, some works paid attention to the quality of the sensing data,
\cite{Whitehouse02,Bychkovskiy03,Feng03,Miluzzo08} but most of them focused on 
simple data applications, such as temperature or relative humidity, since these were the
types of data measured by the first available sensors.
Others \cite{Ramakrishnan11,Stankovic15} used 
synthetic data generated with probability functions, such as Gaussian distribution 
functions, for validating calibration models.
However, the concern for the quality of the data has increased
with the growth of real wireless sensor deployments.
For example, Buonadonna et al. \cite{Buonadonna05} mentioned the 
disappointment when the values of light sensors (calibrated according to the 
manufacturer's specification) in a deployment were compared with the value from a 
calibrated sensor in an environmentally controlled chamber. Ramanathan et
al. \cite{Ramanathan06} described the difficulties in obtaining realistic readings
of ion-selective electrode sensors used for monitoring water quality in a real deployment
in Bangladesh. More recently, Snyder et al. \cite{snyder2013} and Williams \cite{williams2013}
mentioned the lack of knowledge on the performance and long-term reliability of low-cost air pollution sensors.

% some definitions, basic concepts
The classical calibration process mainly consists of calibrating the sensor device 
in a {\it controlled environment}, for example, in a laboratory with high-cost instrumentation,
where the sensor response is measured under different controlled conditions.
When it is not possible to use such a laboratory, or when sensor devices are under 
operation after deployment, the sensor parameters can be self-calibrated and adjusted 
in reference to another sensor of the network, whether calibrated with a
ground-truth\footnote{The {\it ground-truth value} is the reference value obtained
  from a perfectly calibrated sensor. In machine learning, the {\it ground-truth value}
  refers to the accuracy of the training set's  
classification for supervised learning techniques.} \cite{hasenfratz2012fly,spinelle2015} 
reference node, calibrated with respect to already calibrated sensor nodes (e.g., distributed 
calibration), \cite{Tan13} or with respect to not-calibrated sensor nodes (e.g., 
blind calibration) \cite{Balzano07}.

When the sensor response cannot be measured in a controlled environment and the 
sensor parameters have to be adjusted according to other sensor nodes in the 
network, the calibration is said to be in an {\it uncontrolled environment}.
In this case, poor or incomplete calibration can lead to significant errors in 
sensor measurements. Moreover, even if the sensors are calibrated before 
deployment, it is not possible to prevent sensor drift after deployment, especially 
when the lifetime of sensor systems can be as long as years. In some cases, even 
costly high-accuracy sensors produce faulty data when aging. Therefore, it 
is necessary to automatically detect drifts and miscalibration in order to correct sensors' 
measurement after deployment to ensure the trustworthiness of long-term WSNs.

% what is already done
While important, ensuring data quality in WSNs presents several challenges \cite{Alexander16}.
The quantity of data is 
large, but the quality is low, since the data can be subject to many types of
faults. In a real sensor network deployment, Buonadonna et al. \cite{Buonadonna05}
observed that failures can occur in unexpected ways which provide inaccurate data.
Koushanfar et al. \cite{Koushanfar03}
have identified several facets of faulty data and suggested five phases of 
testing sensor-based systems. Ni et al. \cite{Ni09} defined
{\it data faults} to be data reported by a sensor that are inconsistent with the 
phenomenon of interest. Measurement faults can be produced by systematic and 
random errors. Each measurement has unpredictable random errors due to environmental 
noise, the precision of the equipment, or sensor manufacturing defects. 

% What we do
This survey provides insight into existing calibration approaches, models, and methods 
that have been proposed for an efficient
calibration process when expensive equipment for controlled calibration is not available,
or when the sensor nodes are already deployed in the field.
Calibration is a technically challenging task mainly due to the existence of random 
noise and the absence of suitable general models. Calibration techniques and 
models depend on the physical space of the signal measured, on the availability of
ground-truth data, and on the capabilities of the sensor nodes.
Although there is a rich variety of papers related to calibration models
and algorithms for different sensors, we have not found any survey that has collected
the definitions, approaches, and models used for calibrating sensors in WSNs. The closest research
area in which part of this knowledge is collected is sensor faults research,
\cite{Ni09,sharma2010} where calibration is considered one of the many causes of 
wireless sensor node faults.

Our main contributions are the following:
\begin{itemize}
\item We first define calibration in uncontrolled environments, stressing that an uncontrolled
  environment involves calibration in the field and not necessarily in the presence of accurate
  instrumentation. Moreover, we state the main challenges in calibrating low-cost sensors
  in uncontrolled environments in WSNs.
\item We present calibration models and the difficulty in choosing a specific calibration model
  in the calibration process.
\item We state the quality of information (QoI) metric generally used to evaluate the performance
  of a calibration process.
\item We define calibration approaches and attributes in order to calibrate a sensor.
  These include specifying the conditions that the calibration process follows,
  specifically, what the measurement area is, how many sensors are involved in
  the calibration process, what the knowledge of the physical phenomenon is,
  what the position of the calibration sensor is with respect to reference data
  when reference data exist, how many times the calibration process is performed,
  how the information is processed with respect to the
  normal application operations, and where the calibration is performed.
\item Finally, we present guidelines on how to calibrate a sensor network.
\end{itemize}

% paper organization
The aim of this survey, then, is to provide a systematically characterized taxonomy of 
approaches, attributes, and models for the calibration of sensors in WSN in uncontrolled 
environments. Sections~\ref{Sec:Cal_Background}, \ref{Sec:Cal_Model}, and \ref{Sec:Cal_Accuracy}
  present the basics of sensor calibration, calibration models, and
 different validation metrics used to evaluate the accuracy 
 of a calibration model. Section~\ref{Sec:Forms} focuses on the different attributes
 and approaches of a calibration process.
Section~\ref{Sec:CalGuide} gives guidelines in how to calibrate a low-cost sensor network.
Section~\ref{Sec:Open} discusses the existing research
challenges in the field of calibration in uncontrolled environments.
Finally, Section~\ref{Sec:Conclusion} summarizes and concludes the paper.

%
%***************************************************************************/
%                                  CALIBRATION                                                       
%***************************************************************************/
\section{Background on Calibration}
\label{Sec:Cal_Background} 

By definition, calibration refers to the process of correcting 
systematic errors (i.e., biases) in sensor readings often by comparing a known 
measure from a first device with an unknown measure of a second device to adjust the parameters 
that rule this second device, in order to provide 
an accurate measurement. The term has also often been used to refer to the process of 
adjusting the raw sensor readings to obtain corrected values 
by mapping them into standardized units. Traditional single-sensor calibration often 
relies on providing a specific stimulus with a known result, thus creating a direct 
mapping between sensor outputs and expected values. Consequently, such calibration 
for a sensor is often subject to specific ranges and operating-condition restrictions, 
which are reported in the manufacturer's specifications of the sensor. This type of 
calibration can be performed in the factory, during the production stage, manually in the field, or both. 
In addition to component-level calibrations, sensors usually 
must be calibrated at the device level when used as part of a measurement system. 
Moreover, recalibration is usually required in order to ensure proper operation of 
a measurement device, as aging and other factors affect sensors and other measurement 
hardware over time.
In wireless sensor networks, when this process is performed in the field, in the absence of an environmentally 
controlled chamber, we call it 
self-calibration in an uncontrolled environment\footnote{Henceforth, throughout the paper,
whenever we speak of calibration, we refer to self-calibration in uncontrolled environments in
WSNs. We stress that accurate instrumentation can be used in the calibration process
in uncontrolled environments. The difference with respect to controlled environments is that in
controlled environments, the calibration conditions can be manipulated manually (in a controlled chamber) to match
the manufacturer's data-sheet conditions, while in non-controlled environments,
there may be instrumentation that provides reference information but 
depends on conditions that cannot be manipulated.}. 

%=============================================================================================
%                                    CALIBRATION CHALLENGE  
%=============================================================================================
\subsection{Calibration Challenges}
\label{Sec:Cal_Challenges} 
Calibration in sensor networks is challenging because of several reasons  
\cite{Whitehouse02,Bychkovskiy03,Feng03,Buonadonna05}. First, the sensor system network consists of a large 
number of devices typically with no calibration interface. Therefore, in-place sensor calibration schemes 
become impractical, time consuming, and difficult to achieve. Moreover, sensor nodes are 
exposed to environmental noise and hardware failures, and the mismatch between factory 
calibration conditions and in-field conditions makes the calibration of sensors challenging.
In addition, different sensors require different calibration procedures, and the reference 
values might not be readily available. 
So, in general, even when the response function is known, calibration in uncontrolled environments is difficult.
An example is an ozone (O$_3$) sensor whose output value depends not only on 
the O$_3$ concentrations but also on the environmental conditions -- temperature and relative 
humidity -- of the place in which the sensor is deployed. Conducting a multiple linear 
regression with respect to ground-truth O$_3$ concentrations in a place with different 
environmental conditions than those of the place in which the sensor node will be deployed can produce 
large errors in the predicted O$_3$ concentrations \cite{spinelle2015}. Yamamoto et al.
\cite{yamamoto2017} have recently observed a similar behavior in temperature sensors.
Temperature sensors calibrated in a place behave differently when placed in another
location because of the difference in environmental conditions (such as solar radiation,
humidity, wind speed, rainfall, and azimuth) between the two locations.

In large-scale networks, two more challenges add to the inherent difficulties of 
calibrating a sensor: (i) the need to calibrate a massive number of sensors and (ii) the inconveniences 
of physically accessing the sensors, as they may be deployed in far and harsh, or even hostile, 
environments. Additionally, the sensors are presumed to stay active for long periods of time 
after deployment; therefore, they are expected to be checked regularly against standard instruments 
to ensure the measurements' quality and to allow periodic recalibration necessary because of the
loss of accuracy caused by environmental conditions or internal defects. 
However, calibration is expected to minimize systematic and random errors; increase
the accuracy of sensor readings with respect to the reference model; and manage the
aforementioned constraints, or requirements, in addition to the limited capability of
available sensors to provide accurate data at a low cost and without overhead mechanisms.
Hence, a good calibration process should consider the following aspects \cite{Yu15}:
i) time and monetary cost, ii) disruption to normal operation, iii) access to sensors
in places difficult to access, and iv) calibration of a large number of sensors in the field.
For example, in many large cities, sensors are installed in traffic lights or lamp-posts.
Installing these sensors implies stopping vehicle traffic, and access to these sensors is
difficult, costly, and time consuming.
%=============================================================================================
%                                  CALIBRATION ERRORS 
%=============================================================================================
\subsection{Calibration Faults}
\label{Sec:Cal_Errors} 
Sensor error is defined as an unexpected value from the sensor device after its deployment.
This type of error comprises precision degradation, reading bias, drift, noise, 
or sensor failure, which is generally due to miscalibration. The miscalibration error, also known as
\emph{measurement error}, can be defined as the difference between the values indicated by a sensor
device and the true values provided by a reference model. Data faults due to miscalibration can manifest
in different forms by lowering the accuracy of sensor measurement, its precision, or both. Usually,
these errors are referred to as bias, gain, drift, or offset. The \emph{offset, gain, and drift} are the well-known 
miscalibration errors that were first identified in \cite{Bychkovskiy03,Ramanathan06,Balzano07}
and discussed by \cite{Ni09,sharma2010} as the main three different forms of sensor calibration
faults in uncontrolled environments. \emph{Offset} arises when
the measured value $Y$ differs from the true value $X$ by a constant amount $\beta_0$, and it
can be determined by measuring the sensed value when the ground-truth value is zero.
When \emph{offset} error is present, the measured values follow similar patterns to those of the expected
phenomenon. \emph{Gain} refers to the rate or the amount of change of the 
measured value with respect to the change in the underlying ground-truth value. In the
specific case of a linear response function, the measured value is expressed by the following equation: 
\begin{equation}
Y=  f(\beta, X) = \beta_0 + \beta_1 X .
\end{equation} 
Here, the coefficients $\beta_0$ and $\beta_1$ represent the offset and gain parameters, respectively.
The term \emph{bias} refers to deviations that are systematic, not random, for example, because
one can consistently over- or underestimate the measurements by X units. Bias, then, 
can be employed to express the gain, the offset, or both of them in sensor reading. \emph{Drift} 
refers to the change over time of the gain and offset parameters associated with a sensor's original, 
or factory, calibration formulas, that causes the performance of the sensor to deviate from the real
signal of the expected phenomenon. Honicky~\cite{Honicky11} observed that when a sensor is miscalibrated,
its gain drift can often be bounded to a certain percentage range. For example, the gain of electrochemical
sensors can drift no more than 5\% per year, according to the data sheet~\cite{MICROceL}.
Finally, \emph{noise} refers to the random 
errors in sensor measurements that interfere with the calibration process and alter the calibration 
parameters. In calibration, the noise error $\epsilon$ is usually expressed as follows:
\begin{equation}
Y= f(\beta, X) + \epsilon,
\end{equation} 
where random noise often follows a probability distribution, for example, $\epsilon\sim N(0,\sigma^2)$
normal distribution with zero mean and variance $\sigma^2$, and most calibration approaches
exploit such property to recover from random errors.
%=============================================================================================
%                                       Calibration Model 
%=============================================================================================
\section{Calibration Model}
\label{Sec:Cal_Model}
Defining the calibration model is fundamental, yet its choice affects calibration accuracy.
Sensors can be characterized by a specific {\it response function} relating to
the measured parameters\footnote{The input parameters are called {\it predictors, features, independent 
variables}, or {\it variables} in machine learning and statistical learning terminology.}, defined by the 
set $X$, with the output parameter\footnote{The output is also called {\it response} or {\it 
dependent variable} in machine learning and statistical learning terminology.} defined as $Y$, that is $Y$ 
=$f(\beta,X)$, where $\beta$ is the calibration coefficient. The calibration curve is 
obtained by fitting an appropriate response function to a set of experimental data consisting of 
the measured signal relative to a known ground-truth signal. 
Depending on the type of phenomenon being 
measured, the response function associated with the calibration curve may be linear, logarithmic, 
exponential, or any other appropriate mathematical model.
The following are the representative response functions found in sensor literature.

\subsection{Linear Functions}
\label{Sec:LF}
Linear response is the most popular function in which the ground-truth value $y$ is represented 
as a linear function of the measured value $x$, as follows 
\cite{Bychkovskiy03,Buonadonna05,Balzano07}:
\begin{equation}
y\sim f(\beta,x)= \beta_0 + \beta_1 x,
\end{equation}
where the calibration coefficients are offset $\beta_0$ and gain $\beta_1$. 
Temperature sensors\footnote{Temperature sensors in other fields not considered WSN,
such as a temperature sensor in a 3D printer, can have other function responses that are nonlinear.}
 \cite{Bychkovskiy03,Balzano07} are examples of sensors in a WSN that follow
linear function responses. A polynomial response function can also be used when the relationship
between the response $y$ and features $x$ is curvilinear\footnote{Polynomials on x belong
to linear models since these ones are linear in coefficient $\beta_j$.}: 
\begin{equation}
y\sim f(\beta,x) = \beta_0 + \sum_{j=1}^{M}\beta_j x^j. 
\end{equation}
%%%%%%%%%%%%%%%%%%%%%%%%%%%%%%%%%%%%%%%%%%%%%%%%%%%%%%%%%%%%%%%%%%%%%%%%%%
\begin{table*}[t]\centering
\caption{Data models for arrays of gas sensors~\cite{spinelle2015,spinelle2017}.\label{Tab:GasSen}}{
\centering
\scalebox{0.75}{
\begin{tabular}{|l|l|l|l|}\hline\hline
{\textbf{Sensor model}} & {\textbf{Manufacturer}} & {\textbf{Gas}} & {\textbf{Multiple linear model}} \\\hline\hline
%=============================================================================================
O3B4         &  $\alpha$Sense   &  O$_3$  &  Y = $\beta_0 + \beta_1 x_{O_3} + \beta_2 x_{NO_2} + \beta_3 x_{NO_2} \times x_{H_{2}O_{2}}$ \\\hline
O3\_3E1F     &  Citytech        &  O$_3$  &  Y = $\beta_0 + \beta_1 x_{O_3} + \beta_2 x_{NO_2}$    \\\hline
NO2B4, NO2\_3E50 &  $\alpha$Sense, Citytech  &  NO$_2$ &  Y = $\beta_0 + \beta_1 x_{NO_2} + \beta_2 x_{O_3} + \beta_3 x_{T} + \beta_4 x_{RH}$ \\\hline
MICS\_2710   &  SGX-Sensotech   &  NO$_2$ &  Y = $\beta_0 + \beta_1 x_{NO_2} + \beta_2 x_{O_3} + \beta_3 x_{T}$     \\\hline
MICS\_4514   &  SGX-Sensotech   &  NO$_2$ &  Y = $\beta_0 + \beta_1 x_{NO_2} + \beta_2 x_{O_3} + \beta_3 x_{NO} + \beta_4 x_{T}$  \\\hline
CairClip NO2 &  CairPol         &  NO$_2$ &  Y = $\beta_0 + \beta_1 x_{NO_2} + \beta_2 x_{O_3}$      \\\hline
NO\_3E100    &  Citytech        &  NO &  Y = $\beta_0 + \beta_1 x_{NO} + \beta_2 x_{T}  + \beta_3 x_{RH} $      \\\hline
TGS-5042, MICS-4514 & Figaro, e2V & CO &  Y = $\beta_0 + \beta_1 x_{CO} + \beta_2 RH$      \\\hline
Gascard NG, S-100H & EdinburghSensors, ELT Sensors &  CO$_2$ &  Y = $\beta_0 + \beta_1 x_{CO_2} + \beta_2 x_{T} + \beta_3 x_{RH} $ \\\hline\hline
%=============================================================================================
\end{tabular}
}}
\end{table*}
%%%%%%%%%%%%%%%%%%%%%%%%%%%%%%%%%%%%%%%%%%%%%%%%%%%%%%%%%%%%%%%%%%%%%%%%%%

\subsection{Multiple Linear Functions}
\label{Sec:MLF}
In multiple linear response functions, the response variable $y$ depends linearly on a set of parameters, as follows:
\begin{equation}
y\sim f(\beta,x) = \beta_0 + \sum_{j=1}^{M}\beta_j x_j.
\end{equation}

Gas sensors, such as CO, O$_3$, and NO$_2$ sensors, are examples
\cite{hasenfratz2012fly,spinelle2015,spinelle2017,liu2015using,Saukh15,Maag16,Esposito2017,maag2017scan,Barcelo2018,mueller2017}
of sensors that follow multiple linear responses. Some gas sensors (see Table \ref{Tab:GasSen}) are better fitted,
with interactions of several features following a model such as \cite{spinelle2015,spinelle2017}:
\begin{equation}
y\sim f(\beta,x) = \beta_0 + \sum_{j=1}^{K}\beta_j x^{a_{1j}}_1 \dots  x^{a_{Mj}}_M. 
\end{equation}

An example \cite{mueller2017} is an O$_3$ electrochemical sensor whose response depends on
O$_3$ and temperature (T): \begin{multline}
y\sim f(\beta,x) = \beta_0 + \beta_1 x_{O_3}  + \beta_2 x^2_{O_3}  + \beta_3 x_{T}   + \beta_4 x^2_{T} \\
+ \beta_5 x_{O_3} x_{T} + \beta_5 x^2_{O_3} x_{T} + \beta_6 x_{O_3} x^2_{T}. 
\end{multline}
\subsection{Nonlinear Functions}
Acoustic, seismic, and electromagnetic signals \cite{Feng03,Tan13,Wang09,Gao11} attenuate
with the distance, $d$, from the source of the signal. Let us assume that sensor $n$ is $d_n$
meters from the source that transmits a signal with energy $S$. 
The attenuated signal, $y_n$, at the position of sensor $n$ is~\cite{Tan13}
\begin{equation}
y_n \sim f(\beta_n,x) = S f(\beta_n,d_n),
\end{equation}
where $f(\beta_n, d_n)$ is a decreasing function of $d_n$, and $\beta_n$ is a set of parameters
of the signal decay function $f(\cdot)$ at sensor $n$. Some of the signal decay functions used are
illustrated next.
\\\\
\textbf{Power law decay:} The propagation of mechanical waves, such as acoustic and
seismic signals, follows a power law decay~\cite{Tan13}:
\begin{equation}
y_n \sim f(\beta_n,d_n) = S \frac{1}{(d_n/r_n)^{k_n}}, 
\end{equation}
where $\beta_n=\{k_n,r_n\}$; $k_n$ is the decay parameter, and $r_n$ is the reference
distance determined by sensor shape. 
\\\\
\textbf{Exponential decay:} The intensity of light attenuates with the travel distance
and follows an exponential decay~\cite{Tan13}:
\begin{equation}\label{eq:exp_decay}
y_n \sim f(\beta_n,d_n) = S  e^{\lambda_n d_n},
\end{equation}where $\beta_n=\{\lambda_n\}$; $\lambda_n$ is a decaying parameter, 
and it is referred to as the Lambert absorption coefficient.
%%%%%%%%%%%%%%%%%%%%%%%%%%%%%%%%%%%%%%%%%%%%%%%%%%%%%%%%%%%%%%%%%%%%%%%%%%%
\begin{comment}
\begin{figure*}[t]
\centering
\includegraphics[height=0.3\textheight]{captor-17001-s4-resistor.png}
%\hspace*{-7pt}
\includegraphics[height=0.3\textheight]{captor-17001-s4-scatterplot.png}
%%%
\includegraphics[height=0.3\textheight]{captor-17001-s4-MLR-calibrated-withShuff.png}
%\hspace*{-7pt}
\includegraphics[height=0.3\textheight]{captor-17001-s4-SVR-calibrated-withShuff-v1.png}
\caption{H2020 CAPTOR wireless sensor node measuring ozone concentrations: 
  (a) resistor values recorded by the sensor node; 
  (b) scatterplot: x-axis with normalized
  values recorded by the sensor; y-axis with normalized values given by the reference node;
  (c) calibrated using multiple linear regression (MLR);
  (d) calibrated using support-vector regression (SVR).}
\label{Fig:captor-cal}
\end{figure*}
\end{comment}

\begin{figure*}[t]
  \begin{center}
    \subfigure[Resistor values recorded by the sensor node.]{\includegraphics[scale=0.45]{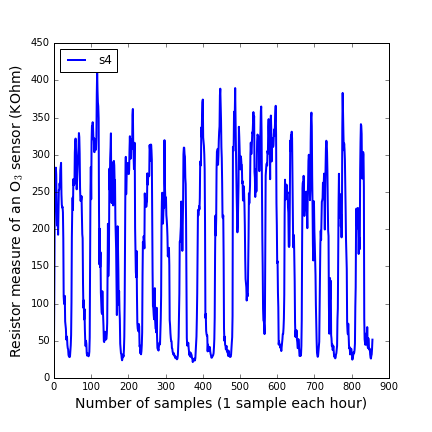}}
    \subfigure[Scatterplot: x-axis with normalized values recorded
  by the sensor; y-axis with normalized values given by the reference node.]{\includegraphics[scale=0.45]{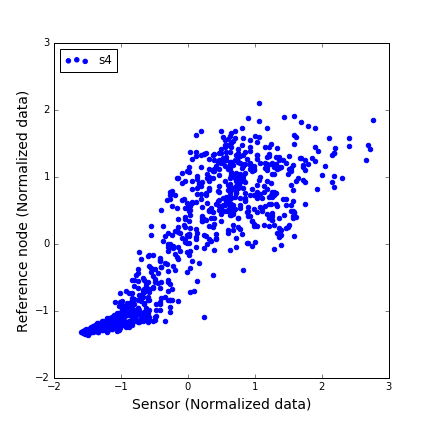}} 
    \subfigure[Calibrated using multiple linear regression (MLR).]{\includegraphics[scale=0.45]{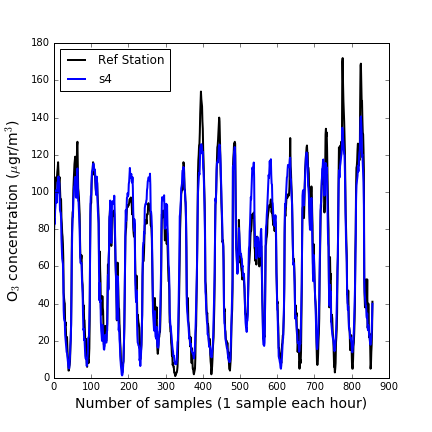}}
    \subfigure[Calibrated using support-vector regression (SVR).]{\includegraphics[scale=0.45]{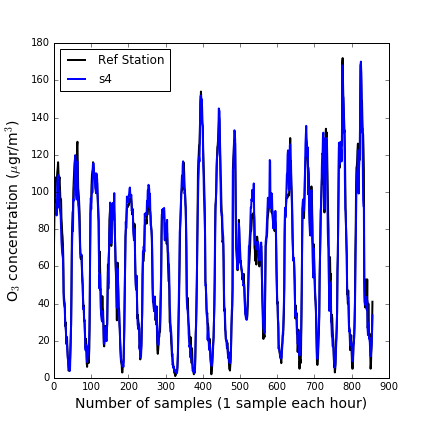}}
  \end{center}
  \caption{H2020 CAPTOR wireless sensor node measuring ozone concentrations.}
  \label{Fig:captor-cal}
\end{figure*}
%%%%%%%%%%%%%%%%%%%%%%%%%%%%%%%%%%%%%%%%%%%%%%%%%%%%%%%%%%%%%%%%%%%%%%%%%%%%%%
Other phenomena are known to yield a linear response. However, when the samples are taken, due
to imperfections of the sensor technology or other causes including the existence of drift or of a maximum
dynamic range, the sensor response is nonlinear. For illustration of this fact, refer to 
Figure~\ref{Fig:captor-cal}a, which shows the samples obtained from a metal-oxide ozone MIC2714 sensor
of a WSN deployed in the H2020 CAPTOR project during summer 2017.
The output of the uncalibrated sensor is in kiloohm. 
The ozone sensor is connected to a voltage divider circuit.
The voltage divider has a load resistor and a variable resistor.
The variable resistor varies according to the ozone concentration, which is what it is obtained at
each measurement. 
The y-axis of Figure~\ref{Fig:captor-cal}a
shows the resistor values recorded at each sample time. The sampling rate is one sample
per hour. The data set comprises approximately 900 samples, collected over 5.5 weeks. The WSN
was collocated with a reference ground-truth node (see Section \ref{Sec:Pos_att}).
Figure~\ref{Fig:captor-cal}b shows the scatterplot of the sensor node data and reference node data.
The values are normalized to their mean and standard deviation. In general, it is well-known 
\cite{spinelle2015,spinelle2017,Maag16} that ozone sensors can be calibrated using a multiple linear
regression using an array of sensors (see Section \ref{Sec:Array} and Table \ref{Tab:GasSen}).
Figure~\ref{Fig:captor-cal}c shows
the calibrated sensor using a frequentist multiple linear regression (MLR). Although the error is
low, $RMSE$ = 9.09 $\mu$gr/m$^3$ with an $R^2$ = 0.947, we can observe from the scatterplot 
(Figure~\ref{Fig:captor-cal}b) that a nonlinear model could improve the calibration error.
The same sensor is calibrated using a nonlinear model called support-vector regression (SVR)
having a $RMSE$ = 6.21 $\mu$gr/m$^3$ with an $R^2$ = 0.976 in Figure~\ref{Fig:captor-cal}d.
This example shows that there is no golden rule for choosing a response function
and that different models will calibrate the same sensor giving different QoI (quality of information),
where QoI is measured in terms of the calibration error given by the model.

Nonlinear responses can be analyzed using a variety of techniques such as
(i) maximum likelihood (ML) \cite{Tan13,Fabeck07} or maximum a posteriori (MAP)
\cite{Moses2002,Ihler04,ihler2005} in its Bayesian version, which finds the
calibration parameters that better fit the data;
(ii) splines \cite{Feng03,Wang09}, which approximate nonlinear response functions by a continuous piecewise linear
function with $m$ segments (i.e., step functions);
(iii) support vector regression (SVR) \cite{Esposito2017,Rossini16}, which classifies data by means
of regression functions that depend on Lagrange multipliers and kernel functions;
(iv) random forest \cite{zimmerman2018}, which constructs an ensemble of decision trees using a training
data set; the mean value from that ensemble of decision trees is then used to predict the value for new input data;
(v) artificial neural networks (ANN) \cite{spinelle2015,spinelle2017,Esposito2017}, which consist
of multiple layers formed by neural units. A single neuron has a number, $I$, of inputs, $x_i$, and one output, $y$.
There are a weight, $w_i$, and a bias, $w_0$, associated with each input. The neuron response is called
{\it activation rule} and captures the nonlinearity of the sensors. 

%%%%%%%%%%%%%%%%%%%%%%%%%%%%%%%%%%%%%%%%  BEGIN TABLE %%%%%%%%%%%%%%%%%%%%%%%%%%%%%%%%%%%%%%%
\begin{table*}[t]\centering
\caption{Simple applications include light, temperature, relative humidity, vibration, and accelerometer sensors.
\label{Tab:LTHR}}
{
\scalebox{0.7}
{
\begin{tabular}{|l||l|l|l|l|}
\hline\hline
%\textcolor{DG}{\textbf{References}} 
%             & \textcolor{DG}{\textbf{Calibration Model}} 
%             & \textcolor{DG}{\textbf{Evaluation metric}}        
%             & \textcolor{DG}{\textbf{Data set}}
%             & \textcolor{DG}{\textbf{Application}}
{\textbf{References}} 
             & {\textbf{Calibration model}} 
             & {\textbf{Evaluation metric}}        
             & {\textbf{Data set}}
             & {\textbf{Application}}
\\\hline\hline
%============================================================================================= 
\cite{Bychkovskiy03}  & Linear regression               &  Mean error             & Real (thermocouple) & Temperature \\
\cite{Feng03}         & Splines/optimization            & Confidence interval     & Real (photovoltaic) & Point-lights \\
\cite{Miluzzo08}      & Distributed consensus           & Mean error & real       & Light   \\
\cite{Buonadonna05}   &  Linear regression              &    Mean error           & Real      & Light, temp., humidity \\
\cite{Balzano07}      & Linear regression               &  Mean \& median error   & Real (thermistor) & Temp., humidity \\
\cite{Wang09}         & Nonlinear/splines              & Mean squared error      & Real      & Light    \\
\cite{Gao11}          & Hidden Markov model             & Recognition accuracy    & Real      & Motion (accelerometer) \\
\cite{Rossini16}      & Support vector regression (SVR) & Mean squared error      & Real      & Light \\
\cite{takruri2010}    & Bayesian                       &  Root mean squared error& Synthetic &  Temperature \\
\cite{Buadhachain13}  & Maximum likelihood              & Absolute error          & Synthetic & Temperature \\
\cite{Kumar13}        & Kriging                         & Root mean squared error & Real      & Temp., humidity, light  \\
\cite{Fujino11,Fujino12,Fujino13}& Gaussian process     & K-L divergence          & Real      & Temperature \\
\cite{Kim2012}        & Linear/nonlinear optimization  & Mean absolute error     & Real      & Vibration (water flow)\\
\cite{Lee14}          & Distributed consensus           & Mean squared error      & Synthetic & Temp., humidity, sound \\
\cite{Wang16}         & PCA + compressive sensing       & Mean squared error      & Real      & Temperature \\
\hline
%=============================================================================================
\end{tabular}
}}
\end{table*}
%%%%%%%%%%%%%%%%%%%%%%%%%%%%%%%%%%%%%%%%%  END TABLE %%%%%%%%%%%%%%%%%%%%%%%%%%%%%%%%%%%%%%%%%
%
%%%%%%%%%%%%%%%%%%%%%%%%%%%%%%%%%%%%%%%%  BEGIN TABLE %%%%%%%%%%%%%%%%%%%%%%%%%%%%%%%%%%%%%%%
\begin{table*}[t]\centering
\caption{Localization, synchronization, and target location applications.
\label{Tab:Local}}
{
\scalebox{0.7}
{
\begin{tabular}{|l||l|l|l|l|}
\hline\hline
{\textbf{References}} 
             & {\textbf{Calibration model}} 
             & {\textbf{Evaluation metric}}        
             & {\textbf{Data set}}
             & {\textbf{Application}}
\\\hline\hline
%=============================================================================================
\cite{Whitehouse02}     & Least squares      &   Mean error  & Real (acoustic) & Localization \\
\cite{Tan13}              & Nonlinear (maximum likelihood) & Detection probability &  Synthetic/real & Target detection \\
\cite{Moses2002}        & Bayesian (maximum a posteriori) & Cram\`er-Rao bound & Real (acoustic) & Localization \\
\cite{Ihler04,ihler2005}& Bayesian (maximum a posteriori) & K-L divergence & Synthetic  & Localization\\
\cite{Fabeck07}         & Nonlinear (maximum likelihood)  & Bayes risk & Synthetic & Target detection \\
\cite{Bolognani10}      & Distributed consensus & Mean, avg. residual & Real (RSSI) & Localization \\
\cite{Stankovic12a,Stankovic15}& Distributed macro-calibration & Mean squared error & Synthetic & Synchronization \\
\cite{Ando05}           & Bayesian (iterative) & Error variance & Synthetic & Synchronization \\
\hline
%=============================================================================================
\end{tabular}
}}
\end{table*}
%%%%%%%%%%%%%%%%%%%%%%%%%%%%%%%%%%%%%%%%%  END TABLE %%%%%%%%%%%%%%%%%%%%%%%%%%%%%%%%%%%%%%%%%
%
%%%%%%%%%%%%%%%%%%%%%%%%%%%%%%%%%%%%%%%%  BEGIN TABLE %%%%%%%%%%%%%%%%%%%%%%%%%%%%%%%%%%%%%%%
\begin{table*}[t]\centering
\caption{Air pollution and water chemistry applications.
\label{Tab:AirPoll}}
{
\scalebox{0.7}
{
\begin{tabular}{|l||l|l|l|l|}
\hline\hline
{\textbf{References}} 
             & {\textbf{Calibration model}} 
             & {\textbf{Evaluation metric}}        
             & {\textbf{Data set}}
             & {\textbf{Application}}
\\\hline\hline
%=============================================================================================
\cite{Ramanathan06}   & Multiple linear regression, splines & Coeff. of determination  & Real & Water chemistry \\
\cite{hasenfratz2012fly}& Multiple linear regression & Mean absolute error & Real &  O$_3$ \\
\cite{spinelle2015,spinelle2017}& Multiple linear regression, neural networks & Mean bias error & Real & O$_3$, CO, CO$_2$,NO, NO$_2$ \\
\cite{liu2015using}       & Bayesian multiple linear regression & Residual sum of squares & Real & CO, CH$_4$, C$_3$H$_8$, CeO$_2$, NiO  \\
\cite{Saukh15}            & Linear and geometric regression & Root mean squared error & Real & O$_3$, CO \\
\cite{Maag16}             & Multiple linear regression & Root mean squared error & Real & O$_3$, CO, NO$_2$ \\
\cite{Esposito2017}       & Multiple linear regression, Gaussian processes, neural networks, & Mean absolute error & Real & O$_3$, NO, NO$_2$ \\
& support vector regression &&&\\
\cite{maag2017scan}       & Optimization  & Mean squared error & Real & O$_3$, CO, NO$_2$ \\
\cite{Barcelo2018}       & Multiple linear regression   & Root mean squared error & Real & O$_3$ \\
\cite{Tolle05}          & Multiple linear regression &    Mean error      & Real & Photosynthetically active radiation (PAR)\\
\cite{Xiang15}            & Maximum likelihood  & Relative bias error & Synthetic & Air pollution \\
\hline
%=============================================================================================
\end{tabular}
}}
\end{table*}
%%%%%%%%%%%%%%%%%%%%%%%%%%%%%%%%%%%%%%%%%  END TABLE %%%%%%%%%%%%%%%%%%%%%%%%%%%%%%%%%%%%%%%%%

\subsection{Choosing a Model}
\label{Sec:Choose-Model}
The choice of calibration model depends on the type of phenomena that the nodes measure,
the resources of the wireless sensor deployment, the type of sensors, and the computation,
storage, and communication capabilities of the sensor node. 
Moreover, the mathematical machinery that has been used in calibration in the past few years is more sophisticated
when the sensor resources increase.
% when the range of sensors developed increases.
More complex calibration approaches
are being formulated with the increased number of wireless sensor network deployments with real applications.
We have to mention that one of the most challenging tasks will be to directly calibrate on a low-cost node.
Most low-cost wireless sensor nodes have too low capabilities to be able to be calibrated on-line.
Light-computational models for low-capability nodes are one of the most difficult challenges that will have
to be solved. For increasing wireless sensor capability nodes, higher computational techniques
can be implemented, facilitating on-line calibration or recalibration mechanisms.
For example, SVR and ANN require high computational resources that most wireless nodes do not
include. In contrast, gradient descent models typically applied in optimization can
be easily implemented in the node to perform on-line calibration in linear or multiple linear regression. 

For instance, temperature sensors (Table \ref{Tab:LTHR}) typically follow linear response functions and have
been analyzed using linear regression \cite{Bychkovskiy03,Buonadonna05} for calibrating the sensors, 
Bayesian inference for modeling drift \cite{takruri2010}, and maximum likelihood \cite{Buadhachain13},
Gaussian processes \cite{Buadhachain13}, or kriging \cite{Kumar13} for state-space modeling applications. 
Light point sensors have mainly been calibrated using splines \cite{Feng03,Wang09}
or support-vector regression \cite{Rossini16} due to the presence of nonlinearities,
and they have been analyzed with distributed consensus \cite{Miluzzo08} to show how light sensing 
is affected by sensor orientation. 
In the case of localization applications (Table \ref{Tab:Local}), the most used calibration
technique is maximum a posteriori \cite{Moses2002,Ihler04,ihler2005} although
distributed consensus \cite{Bolognani10} is also used when spatial redundancy is present.
Target detection \cite{Tan13,Fabeck07} uses maximum likelihood to estimate detection probabilities.
For synchronization applications, Stankovic et al. \cite{Stankovic15,Stankovic12a} defined the distributed
algorithms that converge fast to estimate the offset and gain of the calibration function. 
Ando et al. \cite{Ando05} used an iterative Bayesian algorithm for estimating the offset time
of each sensor after the $k$-$th$ event.

To finalize the set of examples, to monitor air pollution (Table \ref{Tab:AirPoll}), we mainly 
need arrays of sensors, because pollutants depend on several factors (Table \ref{Tab:GasSen}). 
In general, air pollution sensors are analyzed using multiple linear regression (MLR)
\cite{Ramanathan06,hasenfratz2012fly,liu2015using,Maag16,Esposito2017,Barcelo2018,Tolle05}, although
when nonlinearities due to the chemical composition of the sensor appear, techniques such as artificial
neural networks (ANN), support-vector regression (SVR) or random forest are used
\cite{spinelle2015,spinelle2017,Esposito2017,zimmerman2018}.

\subsection{Open Calibration Data Sets}
\label{Sec:DataSets}
There is no centralized repository of open data sets for the calibration of low-cost sensors.
Most of the data sets found on the Internet are data collected by sensors operating on real networks.
For example, CRAWDAD\footnote{https://crawdad.org/}
contains data from sensors such as temperature, accelerometers, location, or RF signal
strength sensors. The major drawback of these repositories is that the data they contain
are from applications and are not data collected in the calibration phase. This makes it difficult
in most cases to use these data to investigate algorithms or calibration mechanisms.

In the field of calibration, some authors have published data obtained
from experiments or deployments designed for the calibration of sensors at WSNs. For example,
the UCI machine learning repository\footnote{https://archive.ics.uci.edu/ml/}
  contains some data sets related to the machine learning
algorithms used in the calibration of low-cost sensors. Examples are the data set published by 
Fonollosa et al. \cite{fonollosa2015chemical} used in their
investigation of calibration of chemical sensors \cite{vergara2012chemical,rodriguez2014calibration}
and the data set published by De Vito et al. \cite{de2009co,de2018} and used in the calibration
of air pollution sensors.
%=============================================================================================
%                                       Accuracy of the model
%=============================================================================================
\section{Accuracy of the Model}
\label{Sec:Cal_Accuracy}
Given the twofold goal of a calibration process -- to recover the true values of a phenomenon and
to detect sensory faults -- a good estimation of the calibration parameters and the accuracy of the process
determine the effectiveness of the model. The way to quantify the accuracy or QoI of a fitting model is
by minimizing an error function~\cite{bishop2006pattern} that measures the misfit between the output
$Y$ and the response function f($\beta$,x) for any given value of $\beta$ and the data set $X$. If the
size of the data set $Y$ is $K$, then a choice of an error function is the square of the errors between
the predictions $f(\beta,x_k)$ for each data point $x_k$ and the target value $y_k$:
\begin{equation}\label{eq:RSS}
E(\beta) = RSS =  \sum^K_{n=1} (f(\beta,x_k)-y_k)^2,
\end{equation}
where RSS stands for residual sum of squares. Then, the goal is to minimize the error function
\begin{equation}
\begin{array}{ll}
    \textrm{min}          & E(\beta) \\
    \textrm{var.}         & \beta
\end{array}.
\end{equation}
The calibration parameters, $\beta^*$, are the solution of the minimization problem.
The error is positive except whenever the function $f(\beta^*,x_k)$ passes exactly 
through each target point $y_k$, in which case the error is zero. The following
QoI metrics can be defined.
\begin{enumerate}[(i)]
\item \emph{The mean squared error (MSE)} measures the average of the squares of the errors. It is the
second moment (about the origin) of the error and thus incorporates the variance of the calibration curve.
\item \emph{The root mean squared error ($RMSE$)} allows comparing different sizes of data sets,
  because it is measured on the same scale as the target value $y_k$.
\item \emph{The coefficient of determination ($R^2$)} measures the proportion of variability in $Y$ that can
be explained using $X$, and it is bounded between $0$ and $1$. A value of $R^2$ close to $1$ indicates that
a large proportion of the variability in the response has been explained by the regression.
\item \emph{The normalized mean bias error ($NMBE$)} represents effectively a total percent error. The
use of percent differences rather than absolute differences normalizes the size of errors in the calibrated 
measurements.
\item \emph{The mean absolute error (MAE)} is a quantity used to measure how close the calibrated
measurements are to the ground-truth data.
\end{enumerate}
Other times, the calibration parameters are represented by a probability
distribution. An example is a response function that follows a normal distribution of mean
$y_k$=$f(\beta,x_k)$ and variance $\sigma^2$. Now, the objective is to obtain a distribution
with mean $\hat{\beta}$ and variance $\hat{\sigma}^2_{\beta}$. In this case, other QoI
metrics can also be obtained.
\begin{enumerate}[(vi)]
\item \emph{The Cram\`er-Rao bound (CRB), or information inequality}, expresses a lower bound on the
variance of a parameter's estimators.
\end{enumerate}
There are metrics that allow measuring the accuracy or precision of the estimation
and depend on the algorithm or application defined by the authors. One example is the
Kullback-Leibler ($KL$) divergence, which measures how one probability distribution
diverges from a second expected probability distribution \cite{ihler2005,Fujino13}.
Other examples are the average value of the relative bias error ($RBE$), the average
value of the relative deviation error ($RDE$) \cite{Xiang15}, and the Bayes risk in
event detection applications \cite{Fabeck07}.

Tables \ref{Tab:LTHR}, \ref{Tab:Local}, and \ref{Tab:AirPoll} show how different sensor applications
use different QoI metrics to assess the performance of the sensor network. The most used
metrics are $MSE$, $RMSE$, and $R^2$. Other authors use the target diagram \cite{Pederzoli12}
that provides information in one plot about the mean bias (MB), the standard deviation, the RMSE,
the centered root mean square error (CRMSE), and the correlation coefficient, R. However, the QoI
metrics in general are linked to the calibration model selected and to
the application. For example, $R^2$ and $RMSE$, $MAE$, and $MSE$ are good metrics to assess whether
the assumptions of the model are correct and estimate whether the error is large. 

%%%%%%%%%%%%%%%%%%%%%%%%%%%%%%%%%%%%%%%%%%%%%%%%%%%%%%%%%%%%%%%%%%%%%%%%%%%%%%%
%============================== Calibration FORMS =============================
%%%%%%%%%%%%%%%%%%%%%%%%%%%%%%%%%%%%%%%%%%%%%%%%%%%%%%%%%%%%%%%%%%%%%%%%%%%%%%%
%%%%%%%%%%%%%%%%%%%%%%%%%%%%%%%%%%%%%%%%%%%%%%%%%%%%%%%%%%%%%%%%%%%%%%%%%%%
\begin{figure*}[t]
\centering
\includegraphics[width=.75\textwidth]{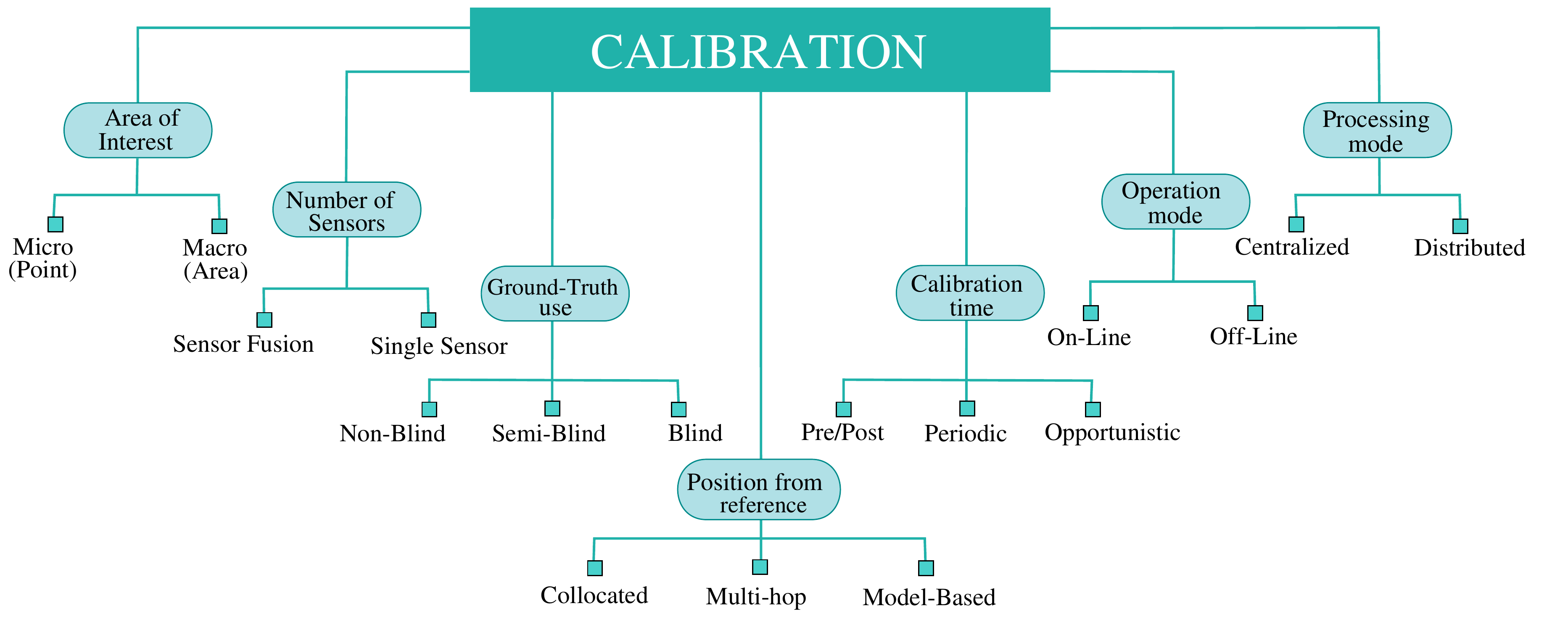}
\caption{Illustration of the calibration attributes that make up a calibration approach.}
\label{Fig:Taxonomy}
%\vspace*{-15pt}
\end{figure*}
%%%%%%%%%%%%%%%%%%%%%%%%%%%%%%%%%%%%%%%%%%%%%%%%%%%%%%%%%%%%%%%%%%%%%%%%%%%
\section{Calibration Approaches and Attributes}
\label{Sec:Forms}
Calibration in sensor networks has been applied to many fields including 
environmental and air quality monitoring \cite{spinelle2015,Saukh15,Xiang15}, 
weather monitoring \cite{Bychkovskiy03,Balzano07,Yu15,Buadhachain13,Fujino11,Lee14,Tolle05},
localization \cite{Whitehouse02,Moses2002,Ihler04,Bolognani10}, synchronization \cite{Stankovic15,Ando05}, 
target discovery \cite{Tan13,Wang09}, robotic, electronic, and radio sensing \cite{LaMarca02,Divi09,Ling15}, and 
water flow monitoring \cite{Kim2012}. These calibration attempts have taken different 
\emph{forms, or approaches,} and have employed a number of calibration \emph{models} (section \ref{Sec:Cal_Model}).
A calibration \emph{approach} is the description of the set of attributes that will characterize the
calibration of the sensor. For example, some authors assume that the node has ground-truth data available,
while others assume that no such data are available. Some calibration models are applied at a centralized
server, while others are distributed. In general, there is a set of \emph{attributes} that capture a specific
calibration approach and that have been used and defined in the literature.
We next describe the attributes of the general characteristics that a calibration approach has.
In order to calibrate a wireless sensor network in uncontrolled
environments, the following questions have to be answered: 
\begin{itemize}
\item  How is the measurement area? Calibration can be {\it micro} (i.e., performed at given points)
  or {\it macro} (i.e., performed in given areas). 
\item  What is the number of sensors involved? Calibration can be {\it single} (i.e., using only one sensor)
  or {\it sensor fusion}, also called {\it multi-sensor data fusion}. Sensor fusion includes the case
  of an {\it array of sensors} (i.e., using multiple sensors). 
\item  What is the knowledge of the physical phenomenon? Calibration can be {\it non-blind}
  (i.e., with full information), {\it semi-blind} (i.e., with partial information), or {\it blind}
  (i.e., with no information). 
\item  What is the position of the uncalibrated nodes with respect to the ground-truth node?
  Calibration can be performed by nodes that are {\it collocated}, {\it multi-hop} (i.e., iterative),
  or {\it model-based}. 
\item  How many times and when is the calibration performed? Calibration can be carried out
  during {\it pre-deployment} (i.e., before the deployment of nodes), {\it post-deployment}
  (i.e., after the deployment of nodes), {\it opportunistically} (i.e., whenever possible),
  or {\it periodically} (i.e., at given intervals of time). 
\item  How is the information processed with respect to the normal operation? Calibration can
  be {\it off-line} (i.e., network not operative, normally related 
  to calibrating with a set of data samples) or {\it on-line} (i.e., network operative, normally
  related to calibrating at each sample arrival).
\item  Where is the calibration performed? The process can be {\it centralized}
  (i.e., localized at a central station) or {\it distributed} (i.e., among nodes). 
\end{itemize}
Answering these questions is fundamental and facilitates the understanding of different
sensor calibration approaches and of the calibration models that have been employed. 
These calibration attributes are described in the following subsections; Figure \ref{Fig:Taxonomy}
shows a map of the calibration attributes.
%---------------------------------------------------------------------------
%                              MICRO/MACRO
%---------------------------------------------------------------------------
\subsection{Calibration Area} 
Sensor nodes are usually deployed in a specific area. Depending on the area of interest,
calibration can be performed at a given point, in which case it is known as micro, or
device-level, calibration. If the objective is a whole area involving a set of sensors
or all of them, the calibration is called macro, or system-level, calibration. 

\begin{figure}[H]
\centering
\includegraphics[width=1.\columnwidth]{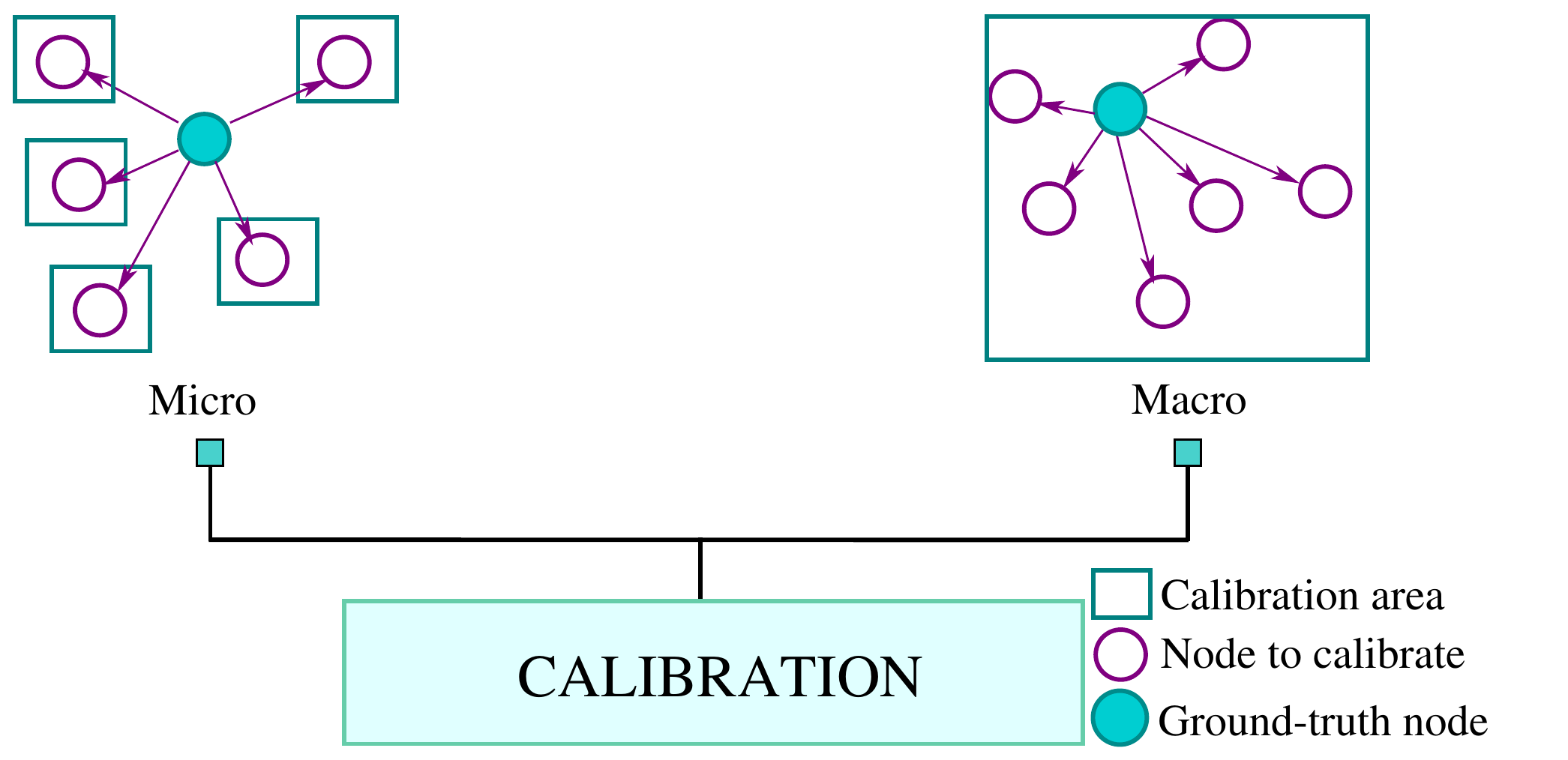}
\caption{Illustration of micro- and macro-calibration in sensor networks.}
\label{Fig:Micro}
\end{figure}

\emph{Micro-calibration} refers to the method that tunes each individual sensor to output
accurate readings at a specific given point (location) of the area, as illustrated in
Figure \ref{Fig:Micro}. Micro-calibration algorithms calibrate every sensor according
to a reference node in order to have accurate sensor measurements of the phenomenon monitored.
Examples of micro-calibration are the calibration of a sensor light at a 
given location~\cite{Moses2002} and the calibration of a NOx sensor \cite{spinelle2015} with
respect to a ground-truth reference station.

\emph{Macro-calibration} assumes a system-wide calibration in the area of interest and focuses
on optimizing the measurements of that area as a whole. The main goal of macro-calibration is
not to adjust sensor calibration according to a reference signal. Instead, calibration algorithms
try to maximize the similarity among the measurements of all sensor nodes in the area. Hence,
most macro-calibration algorithms do not require access to reference measurements
\cite{Stankovic15,Tan13,Buadhachain13}. For example, Stankovic et al. \cite{Stankovic15,Stankovic12a}
proposed a distributed macro-calibration approach based on the generalized consensus problem,
in which all the equivalent sensor gains and offsets in the monitoring area covered by the sensor
network should converge asymptotically to equal values, reducing the overall network error.
Another example is that of \cite{Tan13}, in which a detection mechanism was leveraged for
surveillance applications. Sensors give their readings to a centralized node (a cluster-head),
which computes the calibration coefficients for each sensor node that participates in the process,
with the goal of maximizing the detection probability subject to an upper bound of the false alarm
rate specified by the application.

%%%%%%%%%%%%%%%%%%%%%%%%%%%%%%%%%%%%%%%%%%%%%%%%%%%%%%%%%%%%%%%%%%%%%%%%%%%%

%---------------------------------------------------------------------------
%                              WITH/WITHOUT FUSION
%---------------------------------------------------------------------------

\subsection{Number of Sensors}
\label{Sec:Array}
Many calibration techniques use the measurements from a \emph{single sensor}. This approach is very common
\cite{Bychkovskiy03,Feng03,Miluzzo08,Stankovic15,Buonadonna05,Wang09,ihler2005,takruri2010,Fujino11,Lee14,Wang16,Bolognani10,Ando05}
in sensor network calibration, and these measurements 
are used to establish the calibration model for the sensor and estimate the corresponding parameters. 

In contrast, the goal of \emph{sensor fusion}, or \emph{multi-sensor data fusion}, is to combine
information from two or more data sources into a single one that provides a more accurate description
than that of any of the individual data sources, as illustrated in Figure \ref{Fig:Fusion}. Several
approaches can be considered for sensor fusion. The most common approach is to reduce calibration
errors by jointly considering the measurements of multiple sensors. This technique is called
\emph{array of sensors} and aims to reduce the uncertainty of calibration parameters in the data
model. Arrays of different classes of gas sensors
\cite{spinelle2015,spinelle2017,liu2015using,Maag16,maag2017scan,Barcelo2018} 
have proven useful to qualitatively identify gas species using pattern recognition approaches and
quantitatively determine gas composition based on regression models. An example is
a NO$_2$ sensor that is regressed using NO$_2$, O$_3$, temperature, and relative humidity sensors.
Moreover, arrays or networks with the same class of sensors 
\cite{Barcelo2018} can be represented by a \emph{virtual} value that stands for the set of sensors.
In this case, the set of nodes sensing the same physical phenomenon can use a multivariate model,
a hierarchical Bayes model, or a consensus algorithm to obtain the virtual value representing the
calibrated sensor.
%-%%%%%%%%%%%%%%%%%%%%%%%%%%%%%%%%%%%%%%%%%%%%%%%%%%%%%%%%%%%%%%%%%%%%%%%%%%%
\begin{figure}[htb]
\centering
\includegraphics[width=1.\columnwidth]{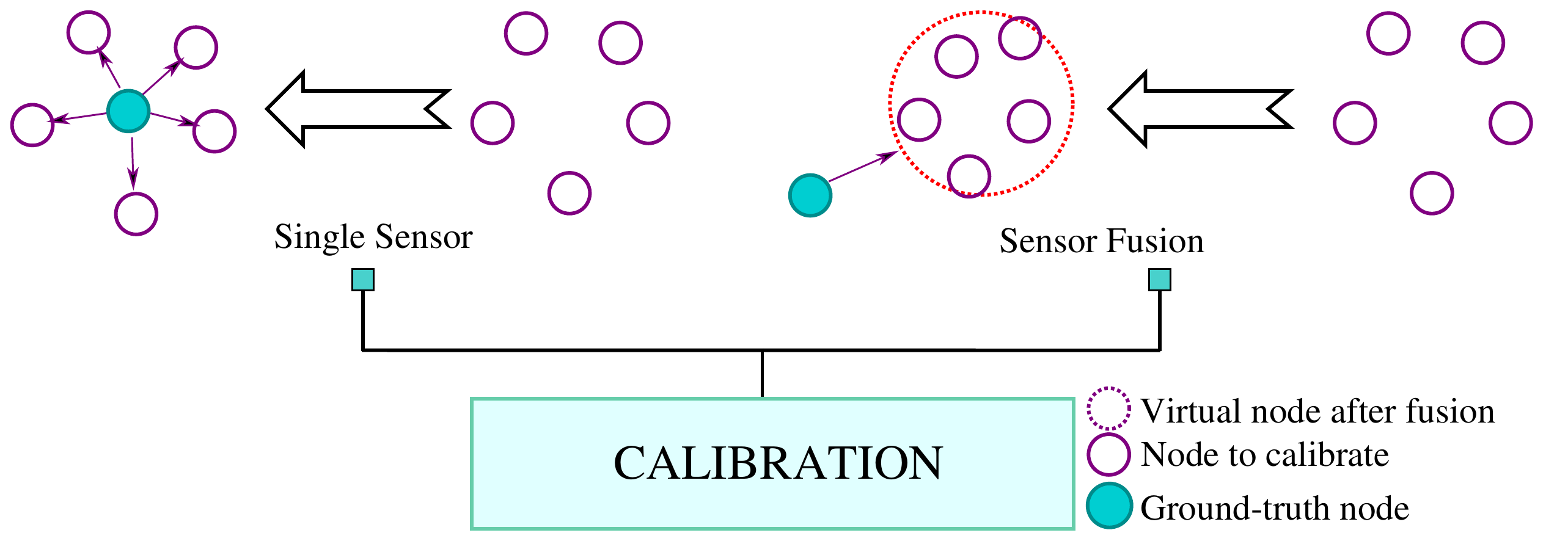}
\caption{Calibration with single sensor and with sensor fusion.}\label{Fig:Fusion-eps-converted-to.pdf}
\end{figure}
%-%%%%%%%%%%%%%%%%%%%%%%%%%%%%%%%%%%%%%%%%%%%%%%%%%%%%%%%%%%%%%%%%%%%%%%%%%%%
Other techniques used for sensor fusion are the following: Tan et al. \cite{Tan13} 
proposed a two-tier system-level calibration of a sensor network. As a first step, each 
sensor learns and transmits its local sensing model to a head node, also called 
fusion-head. Then, the received sensors' measurements are fused in the 
second tier, where a common model is established, and sensors are globally 
calibrated to optimize the system-wide performance. Similarly, Fabeck et al. \cite{Fabeck07} 
considered a network of nodes that send measured data to a centralized node 
according to a binary hypothesis testing problem for detecting the presence of a target. 
The received decisions are combined to yield a final decision. The decision fusion problem
can be viewed as a hypothesis testing problem, with local detection results being the
observations and a Bayes optimal fusion rule 
taking the form of a likelihood ratio test. The minimum probability of error 
associated with the optimal fusion rule is given by a Bayes risk probability 
function. Gao et al. \cite{Gao11} used multi-sensor fusion of four sensors 
attached to the waist, chest, thigh, and side of the body for activity recognition.
They relied on Bayesian techniques to increase the system estimation and achieved between
$70.88\%$ and $97.66\%$ accuracy. 

%---------------------------------------------------------------------------
%                              BLIND/SEMI/NON-BLIND
%---------------------------------------------------------------------------
\subsection{Knowledge of Ground-Truth Data}
\begin{figure}[htb]
\centering
\includegraphics[width=1.\columnwidth]{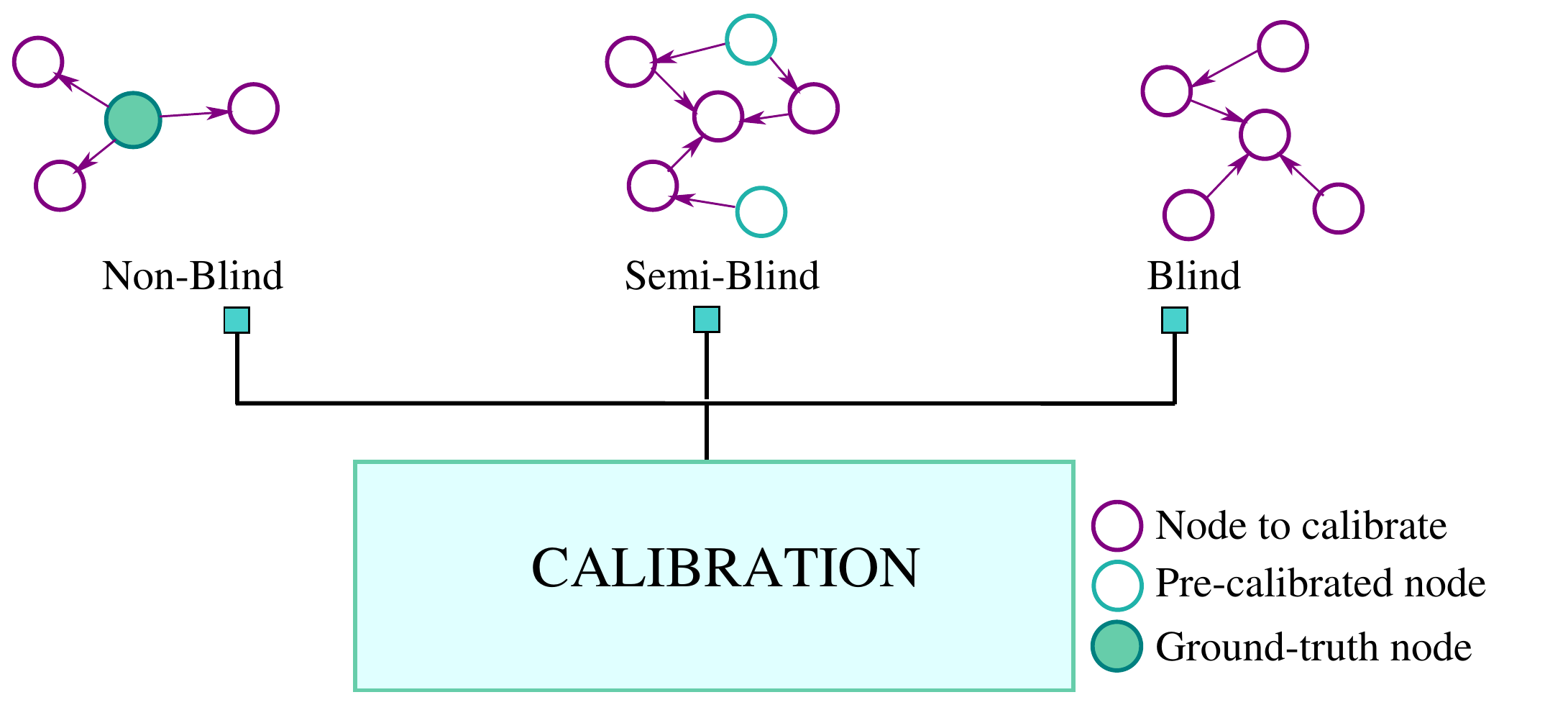}
\caption{Illustration of blind, semi-blind, and non-blind calibration.}
\label{Fig:Blind}
\end{figure}

Calibration can be blind, semi-blind, or non-blind depending on whether or 
not sensory data are processed in the presence of controlled stimuli, a reference 
model, or high-fidelity ground-truth values (Figure \ref{Fig:Blind}). 

%-%%%%%%%%%%%%%%%%%%%%%%%%%%%%%%%%%%%%%%%%%%%%%%%%%%%%%%%%%%%%%%%%%%%%%%%%%%%   
%-%%%%%%%%%%%%%%%%%%%%%%%%%%%%%%%%%%%%%%%%%%%%%%%%%%%%%%%%%%%%%%%%%%%%%%%%%%%

In \emph{non-blind calibration}, also known as \emph{reference-based}, sensors are calibrated
leveraging some reference information. These approaches are based on a priori knowledge of
\emph{ground-truth} data, from high-quality sensors, which can be useful for
comparing the response with the expected values. Accordingly, the calibration parameters are
adjusted using these known data inputs. A good example in which non-blind calibration techniques
are easy to apply is temperature and gas sensors. In the case of gas sensors, in many countries
there are organizations\footnote{The European Environment Agency
  (EEA, http://www.eea.europa.eu/data-and-maps/data/airbase-the-european-air-quality-database-7)
 and the United States Environmental Protection Agency (EPA, https://www.epa.gov/outdoor-air-quality-data)
 publish air pollution ground-truth data from reference stations of many countries in Europe and in the US,
 respectively.} that deploy high-accuracy reference stations to measure air pollution.
These reference stations are not densely deployed due to their high costs 
(approximately 100 thousands euros)\footnote{Similar cost in US dollars.}, 
but they can be used to calibrate low-cost sensors located behind them
(pair-wise or collocated calibration; see Section \ref{Sec:Pos_att}).
Non-blind calibration was applied for sensor network calibration by
\cite{Ramanathan06,hasenfratz2012fly,spinelle2017,Maag16,Fabeck07,Kim2012,Tolle05}.
Moreover, some researchers use commercial instruments to provide ground-truth measurements
that are later used to evaluate calibration models \cite{spinelle2015,spinelle2017}.
These instruments are not calibration chambers in which the conditions can be manually
adjusted; instead, they are accurate devices that allow comparing the low-cost readings
against reference data. 

\emph{Blind calibration} is the approach that calibrates sensor
networks without relying on any controlled stimuli or high-fidelity ground-truth data 
\cite{Balzano07}. Instead, blind calibration schemes leverage signal processing theory,
a priori knowledge, such as physical models and constraints, of the sensed phenomenon, and
exploit the advantages of dense sensor deployments, such as redundant measurements and temporal
or spatial correlation among groups of sensors, to determine the fitting function and estimate
the calibration parameters. Many blind calibration works have been proposed for sensor networks
employing different techniques 
\cite{Ramakrishnan11,Stankovic15,Balzano07,Wang09,Lee14,Wang16,Lipor14,bilen2014convex,gribonval2012blind,schulke2013blind}. 
Balzano et al. \cite{Balzano07} used least squares to solve the linear blind calibration problem;
later, Lipor and Balzano \cite{Lipor14} also proposed blind calibration with model mismatch in which
the subspace measured is not perfectly known, solving the problem using total least squares. 
Ramakrishnan et al. \cite{Ramakrishnan11} estimated the signal signature and simultaneously calibrated 
the sensor nodes using gossip-based algorithms. Stankovic et al. \cite{Stankovic15} proposed a 
distributed algorithm for blind macro-calibration of large sensor networks. Wang et al. \cite{Wang09}
proposed a density guided blind calibration (DGC) scheme for nonlinear mobile sensors by approximating the 
nonlinear calibration functions using piecewise linear functions. DGC relies on the observation 
that different sensors that move in the same region record similar field statistics. 
Other authors~\cite{Wang16} proposed a learning phase in which principal component analysis (PCA)
is used on a subspace of drift-free samples, followed by a calibration phase in which compressed
sensing is used to estimate the sensor drift. Compressed sensing (CS) exploits the fact that the
signal is M-sparse, meaning that only M$<$K out of the K components are non-zero. The authors of 
\cite{bilen2014convex,gribonval2012blind,schulke2013blind} also applied compressed sensing 
to blind calibration frameworks with unknown gains.

\emph{Semi-blind calibration} is an intermediate case between blind and non-blind calibration. 
In semi-blind calibration, a sensor requires partial ground-truth information sent from a subset
of nodes to calibrate itself \cite{Tan13,maag2017scan,Moses2002,Rossini16,takruri2010,Kumar13,Ando05,Xiang15}
or when interacting
with a calibrated sensor \cite{Miluzzo08,Akcan13}, as illustrated in Figure \ref{Fig:Blind}.
For instance, in the semi-blind calibration proposed by \cite{Miluzzo08}, uncalibrated nodes
interact with calibrated nodes and run an average consensus algorithm using their measurements
as partial ground-truth for calibration. Maag et al. \cite{maag2017scan} proposed a multi-hop
calibration scheme in which sensors are calibrated using already calibrated sensors.

%---------------------------------------------------------------------------
%                              COLLOCATED/MULTI-HOP/MODEL-BASED
%---------------------------------------------------------------------------
\subsection{Position from Ground-Truth Nodes}\label{Sec:Pos_att}
\begin{figure}[htb]
\centering
\includegraphics[width=1.\columnwidth]{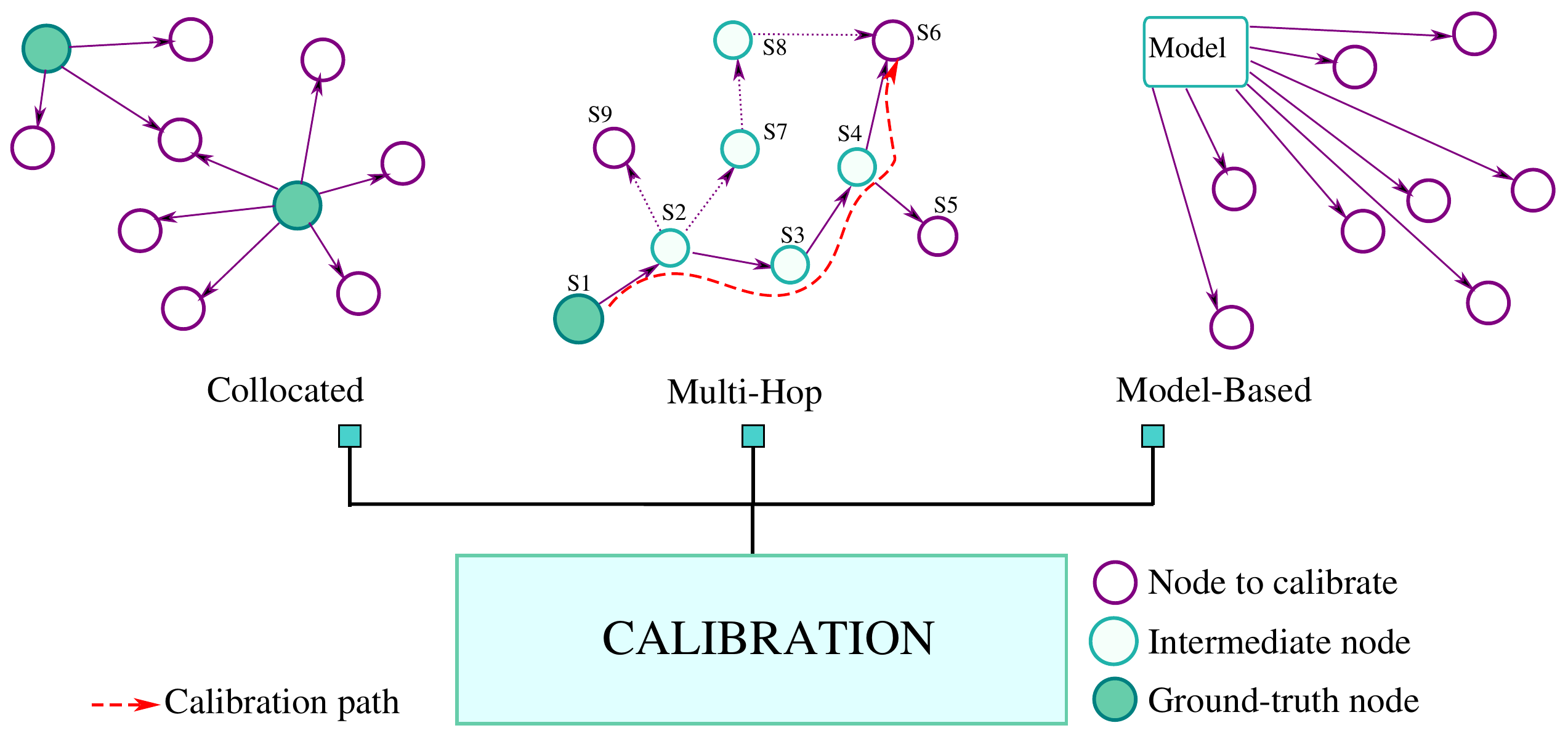}
\caption{Illustration of collocated calibration, iterative, or multi-hop, 
calibration, and model-based.}
\label{Fig:Multi-Hop}
\end{figure}

Depending on the relative position of the uncalibrated node with respect to the reference 
model, calibration can be performed between collocated nodes, 
iteratively hop-by-hop (multi-hop) from the ground-truth, or with the 
local presence of a reference model (model-based), as illustrated in 
Figure \ref{Fig:Multi-Hop}. Following the definition of Saukh et al. \cite{Saukh15},
a process of interest $P$ is a continuous measurable signal
$\eta$:$T\times L\rightarrow D$ with domain $T\subseteq\mathbb{R}^+$ 
and $L\subseteq\mathbb{R}\times\mathbb{R}$, where $T$ represents time and $L$ location.
Let us assume two sensors $s_1$ and $s_2$ that take samples at two times and locations such as ($t$,$l$)
and ($t$',$l$'); then, the process $P$ is bounded in time and space:
\begin{equation}
\mid\eta(t,l)-\eta(t',l')\mid \leq \gamma(\mid t-t'\mid,\mid l-l'\mid),
\end{equation}
with $\gamma$ being a monotonically non-decreasing function. As the authors note, the slower
the function $\gamma$ grows, the more correlated will be the data of the two
sensors.

Saukh et al. \cite{Saukh15} defined a \emph{rendezvous} as the set of spatially and temporally
close pairs of measurements $\Phi^{s1,s2}$ between sensors $s_1$ and $s_2$ within
a time interval $T$ as
%\begin{equation}
\begin{multline}
\label{Eq:Shauk_cond}
  \Phi^{s_1,s_2} =\{ (x_{s1}(t_i,l_i), x_{s2}(t_j,l_j)) \mid (t_i,t_j\in T) \wedge (\mid t_i-t_j\mid 
  \leq \bigtriangleup t)  \\ \wedge(\mid l_i-l_j\mid\leq \bigtriangleup d)\},
\end{multline}
%\end{equation}
where \{$x_{s_1}$\}$\in D_{s_1}$, \{$x_{s_2}$\}$\in D_{s_2}$, and $\bigtriangleup t$ and $\bigtriangleup d$
are temporal and spatial constraints on the required closeness of measurements, respectively.
As it can be observed, the rendezvous defines how close two sensors have to be
to calibrate the uncalibrated sensor with respect to the reference sensor and
the temporal synchronization level of the samples from both sensors.

\emph{Collocated calibration} refers to the situation 
where a node is calibrated using interaction with a close neighbor node 
(collocated) that can sense the same phenomenon and provide reference 
measurements. In terms of the process $P$, collocated calibration involves a sensor
at place ($t$,$l$) that gives ground-truth values, a sensor
at place ($t$',$l$') that is going to be calibrated, and equation (\ref{Eq:Shauk_cond})
to be fulfilled with closed temporal and spatial constraints.
This approach, then, is associated with non-blind calibration
where a node is exposed to the true values available from a reference
ground-truth node \cite{Buonadonna05,spinelle2015,Yu15,Maag16,spinelle2017,Kim2012}.

%-%%%%%%%%%%%%%%%%%%%%%%%%%%%%%%%%%%%%%%%%%%%%%%%%%%%%%%%%%%%%%%%%%%%%%%%%%%%
%-%%%%%%%%%%%%%%%%%%%%%%%%%%%%%%%%%%%%%%%%%%%%%%%%%%%%%%%%%%%%%%%%%%%%%%%%%%%
In \emph{iterative, or multi-hop, calibration}, a node is calibrated using a set of nodes
already calibrated, which are not necessarily ground-truth nodes. The name of multi-hop comes
from the case in which the uncalibrated sensor is $h$-hops from the ground-truth
node, where each hop has iteratively been calibrated.
For illustrating the process, let us define the {\it calibration graph} (Figure \ref{Fig:Multi-Hop})
as a directed graph $G$=$(V, E)$, where $V$=$\{1,2,...,N\}$ is the set of nodes, and
$E$$\subseteq$$V\times$$V$ is the set of edges. The edge $e_{ij}$ between sensor node $s_i$
and sensor node $s_j$ means that $s_j$ is calibrated using node $s_i$ as reference 
node. In this case, an uncalibrated sensor can be calibrated using an $h$-hop path. 
An example (Figure \ref{Fig:Multi-Hop}) is a path
$s_1\rightarrow s_2\rightarrow s_3\rightarrow s_4\rightarrow s_6$,
with $s_1$ being a ground-truth node. Then, $s_2$ would be calibrated using node $s_1$,
node $s_3$ using node $s_2$, $s_4$ using node $s_3$, and $s_6$ using node $s_4$.

Collocated calibration would be the case of having a $0$-hop calibration. Another
special case is when $h$=$1$, called {\it single-hop}. In this situation, a node is calibrated
using already existing neighboring calibrated sensor nodes\footnote{A \emph{pair-wise} calibration is also
  used in this context. Pair-wise implies a rendezvous between two sensor nodes in the calibration.
  If one of them is a ground-truth node, we call the pair-wise a $0$-hop, or collocated, calibration.
  If one of them is an already calibrated node, we call the pair-wise a $1$-hop, or single-hop, calibration.
  It was common in the early literature on calibration in which multi-hop was not well defined
  to call pair-wise any rendezvous, regardless of whether it was an already calibrated node or a
  ground-truth node.}.
In a multi-hop calibration approach, there can be many paths to calibrate an uncalibrated sensor.
Thus, the objective is to minimize the errors committed as a result of the calibration
process~\cite{Bychkovskiy03,Saukh15,maag2017scan}. 

Different paths contribute differently to the post-calibration
{\it skew} of the sensors, where skew is defined as the difference between the calibrated
value of sensor $j$ and the actual value~\cite{Akcan13}. The goal in multi-hop, or iterative, calibration
is to study how the errors propagate along the network path and find calibration paths with minimum skews.
The skew depends on the
length of the paths and on the individual calibration errors of the sensors that the path traverses.
As an example, if sensor $s_1$ is calibrated against a reference sensor and produces measured
data in the range [$x_1$-$\epsilon_1$,$x_1$+$\epsilon_1$] and then a sensor $s_2$ is calibrated
against $s_1$, reporting measured data in the range [$x_2$-$\epsilon_2$,$x_2$+$\epsilon_2$],
the post-calibration skew of sensor $s_2$ will be at most [$\epsilon_1$+$\epsilon_2$]. Akan 
\cite{Akcan13} proved that the skew is at most the sum of the absolute errors along the calibrating path.

In a more recent work, Saukh et al. \cite{Saukh15} investigated whether linear fitting models
suppress multi-hop error propagation in a large-scale mobile sensor network.
Saukh et al. \cite{Saukh15} showed that ordinary least
squares regression suffers from \emph{regression attenuation} or \emph{regression dilution}~\cite{frost2000}.
Regression dilution is the biasing of the regression slope towards zero
caused by errors in the independent variable. This effect increases as the number
of hops increases. The reason is that at every hop, the independent variable $x$ of
the linear fitting adds an error to the model. For example, let us have a path 
$s_1$$\rightarrow$$s_2$$\rightarrow$$s_3$, where $s_1$ is a ground-truth node; then,
assuming a simple linear model,
\begin{equation}
  \begin{cases}
   x_1 \sim y_2 = &  \beta_{02} + \beta_{12}x_2 + \epsilon_2 \\
   y_2 \sim y_3 = &  \beta_{03} + \beta_{13}x_3 + \epsilon_3
  \end{cases},
\end{equation}
with $\epsilon_2$ and $\epsilon_3$ normal distributed errors, $x_{1}$ is the ground-truth data, and 
$x_{2}$ and $x_{3}$ are respectively the measured sensor data and are defined according to
equation (\ref{Eq:Shauk_cond}) in subsection \ref{Sec:Pos_att}. In this case, at each hop,
the calibration coefficient is decreased proportionally by the variance of the error given by that hop 
(regression dilution)~\cite{frost2000}.
The solution is compensating the bias in the slope estimation if the variance of the error of the sensor
node is known. The authors of \cite{Saukh15} mentioned the difficulty of this estimation due to drift
in real sensors
and changing environmental conditions. As an alternative solution, \cite{Saukh15} proposed to use
other fitting techniques that do not suffer from regression dilution, such as geometric mean regression
(GMR). They tested the multi-hop calibration in a real network of O$_3$ and CO sensors, showing that 
ordinary least regression (OLR) produces regression dilution, that 
GMR has very low hop-by-hop error propagation, and that the calibration quality is independent of the 
calibration error of its calibration parent.  Maag et al. \cite{maag2017scan} proposed sensor array
network calibration (SCAN), a multi-hop micro-calibration scheme for mobile sensor arrays.
SCAN minimizes the accumulated error over multiple hops of calibrated sensors. SCAN formulates a
constrained least squares regression optimization problem reducing, and even eliminating under certain
conditions, the regression dilution effect.

In \emph{model-based calibration}, 
a node is calibrated using a set of ground-truth nodes that are not in the 
vicinity of the non-calibrated node. In this case, the ground-truth values allow 
by means of a mathematical model to produce a reference value at the position of 
the non-calibrated node. The values produced by this model now act as ground-truth 
reference data for the estimation of the calibration
parameters~\cite{Feng03,Gao11, Moses2002,Buadhachain13,Kumar13}.
The accuracy of model-based calibration is less than that of the collocated calibration,
since the values estimated in the non-calibrated node are produced by a model that
introduces an error, even if this model uses ground-truth data. 
An example is a source of light. The intensity of the light at source location $d$ can be obtained
by an exponential decaying function \cite{Feng03} that depends on the intensity
of the light source and distance (see equation (\ref{eq:exp_decay}). In another
example, Moses et al. ~\cite{Moses2002} considered the problem of location and orientation of sensors
by measuring the time of arrival (TOA) and direction of arrival (DOA) of the signal emitted
by that source. A third example is given by Whitehouse et al.~\cite{Whitehouse02}, who used received signal
strength information (RSSI) in fusion with acoustic time of flight (TOF) as a method of
auto-calibration for localization services. Some distances between sensors are known a priori,
and this allows taking advantage of anchor nodes or pre-calibrated nodes.

%---------------------------------------------------------------------------
%                              PRE-DEPLOYMENT/PRE-POST-DEPLOYMENT/PERIODIC 
%---------------------------------------------------------------------------
\subsection{Calibration Time and Frequency}
In most cases, the calibration process is performed before the nodes 
are deployed, \emph{pre-deployment calibration} \cite{Feng03,spinelle2015,Wang09,Kumar13}
among others. However, there are situations 
in which the sensors are replaced or recalibrated with certain periodicity. Examples 
are ozone (O$_3$) campaigns that occur in summer in Europe. After a summer campaign, a 
recalibration of the sensors can give information on how the sensors have aged 
and how good the results are. This kind of calibration is called 
\emph{pre-post-deployment} and has two phases: a first calibration before the 
sensors are deployed and a second phase after the deployment. If the sensors 
have aged, the data from the first half of the campaign are predicted with the pre-calibrated 
parameters, while the data from the second part of the campaign are post-calibrated. 

\begin{figure}[htb]
\centering
\includegraphics[width=1.\columnwidth]{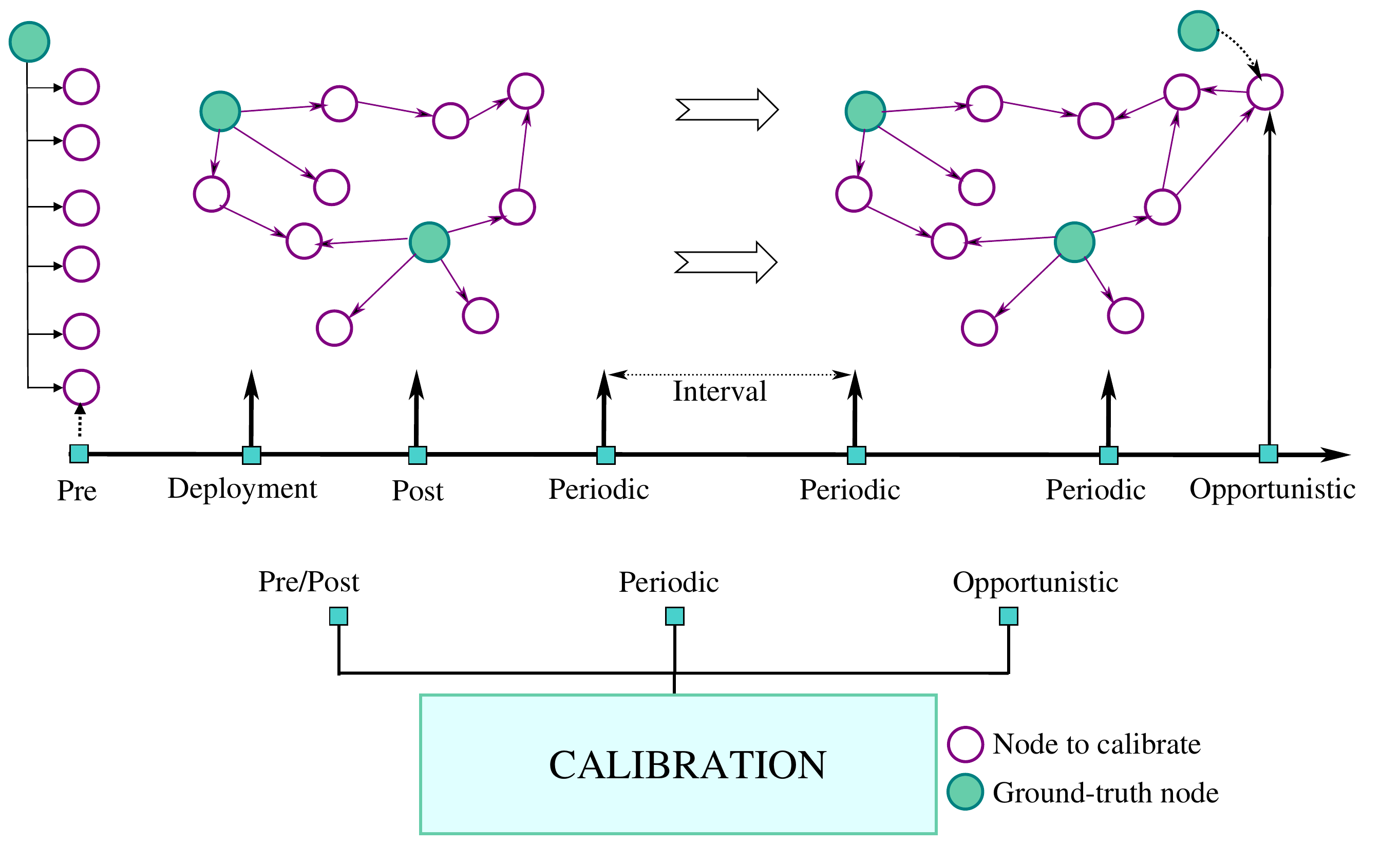}
\caption{Illustration of pre- and post-deployment calibration, periodic 
calibration, and opportunistic calibration.}
\label{Fig:Periodic}
\end{figure}

\emph{Periodic calibration} occurs when the nodes are recalibrated at given time intervals
(see Figure \ref{Fig:Periodic}). The authors of \cite{Saukh15,Maag16,maag2017scan} used two
stations of the local governmental measurement network (​NABEL and ​OstLuft stations) located 
in the city center of Zurich as their high-quality measurements to calibrate low-cost 
sensors deployed on top of public streetcars. Given the known itinerary of public
transportation, the mobile sensors on top of the streetcars are periodically calibrated whenever 
they enter in contact with reference stations that provide ground-truth reference data. 
Characterized by a meeting point with a distance of $50$ meters and a time of $5$ minutes, the 
rendezvous calibration graph of the ozone measurements comprises 500 meeting points between the 
reference stations and mobile sensors within an interval of $10$ days.
%-%%%%%%%%%%%%%%%%%%%%%%%%%%%%%%%%%%%%%%%%%%%%%%%%%%%%%%%%%%%%%%%%%%%%%%%%%%%
%-%%%%%%%%%%%%%%%%%%%%%%%%%%%%%%%%%%%%%%%%%%%%%%%%%%%%%%%%%%%%%%%%%%%%%%%%%%%

\emph{Opportunistic calibration} is used whenever the mobility patterns of sensors are 
unknown and the calibration of sensor measurements relies on the possible rendezvous
with a ground-truth or freshly calibrated node. As an example, Miluzzo et al. \cite{Miluzzo08}
exploited the opportunistic encounter with ground-truth, or calibrated, sensors to 
calibrate uncalibrated mobile sensors. The convergence time of the approach depends 
on the mobility patterns of mobile sensors and on the density of the ground-truth 
nodes.

%---------------------------------------------------------------------------
%                              OFF-LINE/ON-LINE
%---------------------------------------------------------------------------
\subsection{Mode of Operation}
\label{Sec:MO}
\begin{figure}[htb]
\centering
\includegraphics[width=1.\columnwidth]{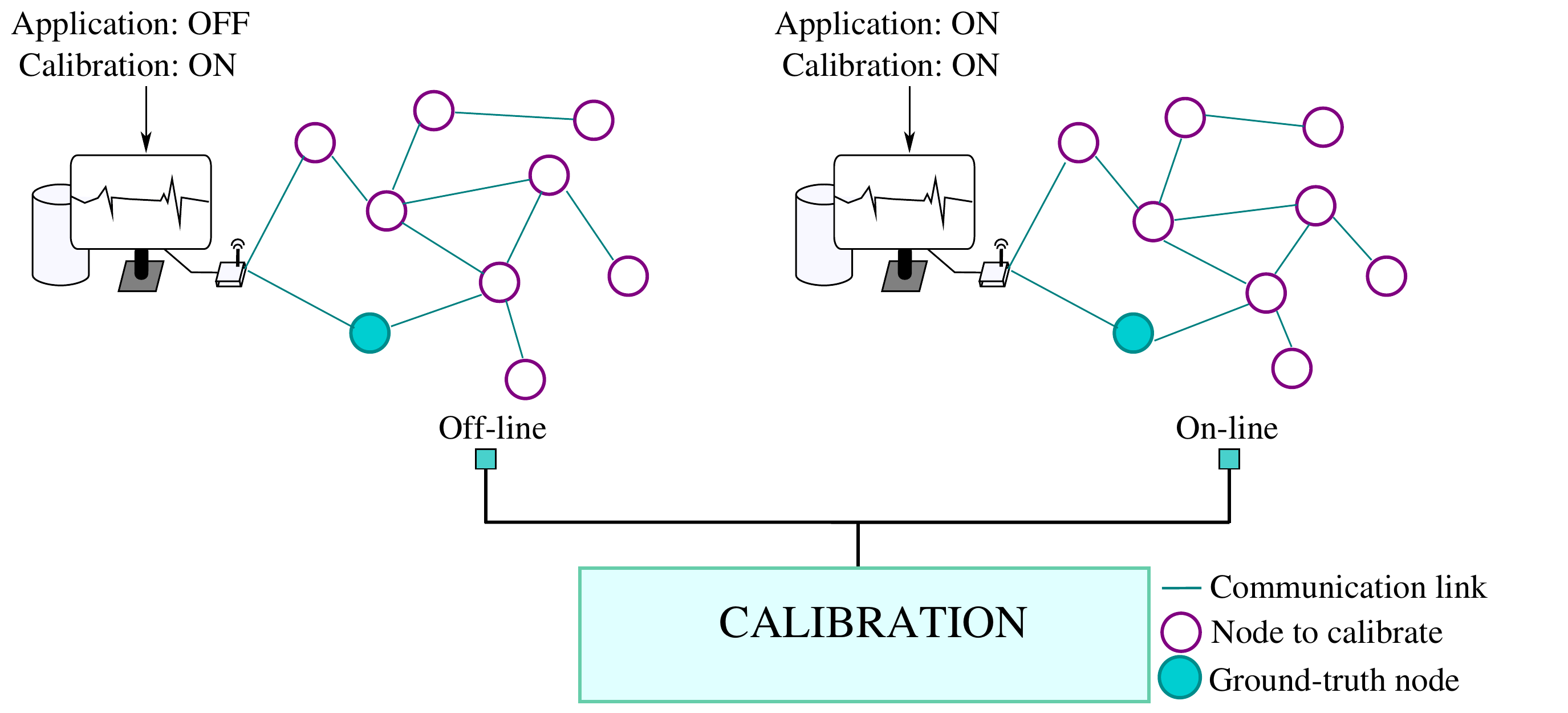}
\caption{Illustration of on-line and off-line calibration in sensor networks.}
%\vspace{-2em}
\label{Fig:Online}
\end{figure}

\emph{Off-line calibration} involves performing sensor calibration when the 
device is not under operation, as illustrated in Figure \ref{Fig:Online}. 
A sensor is under operation when it is taking data that an application is using.
Ramanathan et al. \cite{Ramanathan06} discussed ways to increase the quantity and 
quality of data on-line during a deployment. In off-line calibration\footnote{We
  define off-line and on-line based on ideas presented in the
paper \cite{Ramanathan06}. In this paper, Ramanathan et al. call on-line calibration the case
when in-field users can detect and compensate for problems as they occur. 
We extend the definition to the general case of sensors that are operative.
Some previous papers, such as Feng et al. \cite{Feng03}, named on-line
calibration what was later called blind-calibration in the literature.}, 
a node takes probe data to calibrate the sensor.
In general, off-line calibration requires that the sensor send the data
to a node with enough storage and computation capabilities to
perform the calibration process. Calibration, then, can be carried out using a set
of measurements that are recorded in a repository and later used along with some
reference signal and a known model to calibrate the device.
The goal is to obtain a function that maps the raw data 
recorded by the low-cost sensor to the real value with a minimum prediction 
error. Examples of off-line calibration can be found in 
\cite{Ramanathan06,spinelle2015,spinelle2017,liu2015using,Barcelo2018,Gao11,Moses2002,Kim2012}. 

%-%%%%%%%%%%%%%%%%%%%%%%%%%%%%%%%%%%%%%%%%%%%%%%%%%%%%%%%%%%%%%%%%%%%%%%%%%%%
%-%%%%%%%%%%%%%%%%%%%%%%%%%%%%%%%%%%%%%%%%%%%%%%%%%%%%%%%%%%%%%%%%%%%%%%%%%%%
\emph{On-line calibration} enables performing the calibration 
process during the normal operation of sensors \cite{Miluzzo08,hasenfratz2012fly,Saukh15,ihler2005,Rossini16,takruri2010,Buadhachain13,Lee14}.
For example, in the on-line temperature sensor calibration approach in \cite{Buadhachain13},
sensors cooperate using a gossip-based algorithm and run the expectation-maximization algorithm
that enables them to converge at runtime on local aggregate means in order to compute a set of
calibration parameters. In general, single-sensor on-line calibration requires communication with the
ground-truth node or among the nodes that participate in the calibration process.
%---------------------------------------------------------------------------
%                              Centralized/Distributed
%---------------------------------------------------------------------------
%-%%%%%%%%%%%%%%%%%%%%%%%%%%%%%%%%%%%%%%%%%%%%%%%%%%%%%%%%%%%%%%%%%%%%%%%%%%%
\begin{figure}[htb]
\centering
\includegraphics[width=1.\columnwidth]{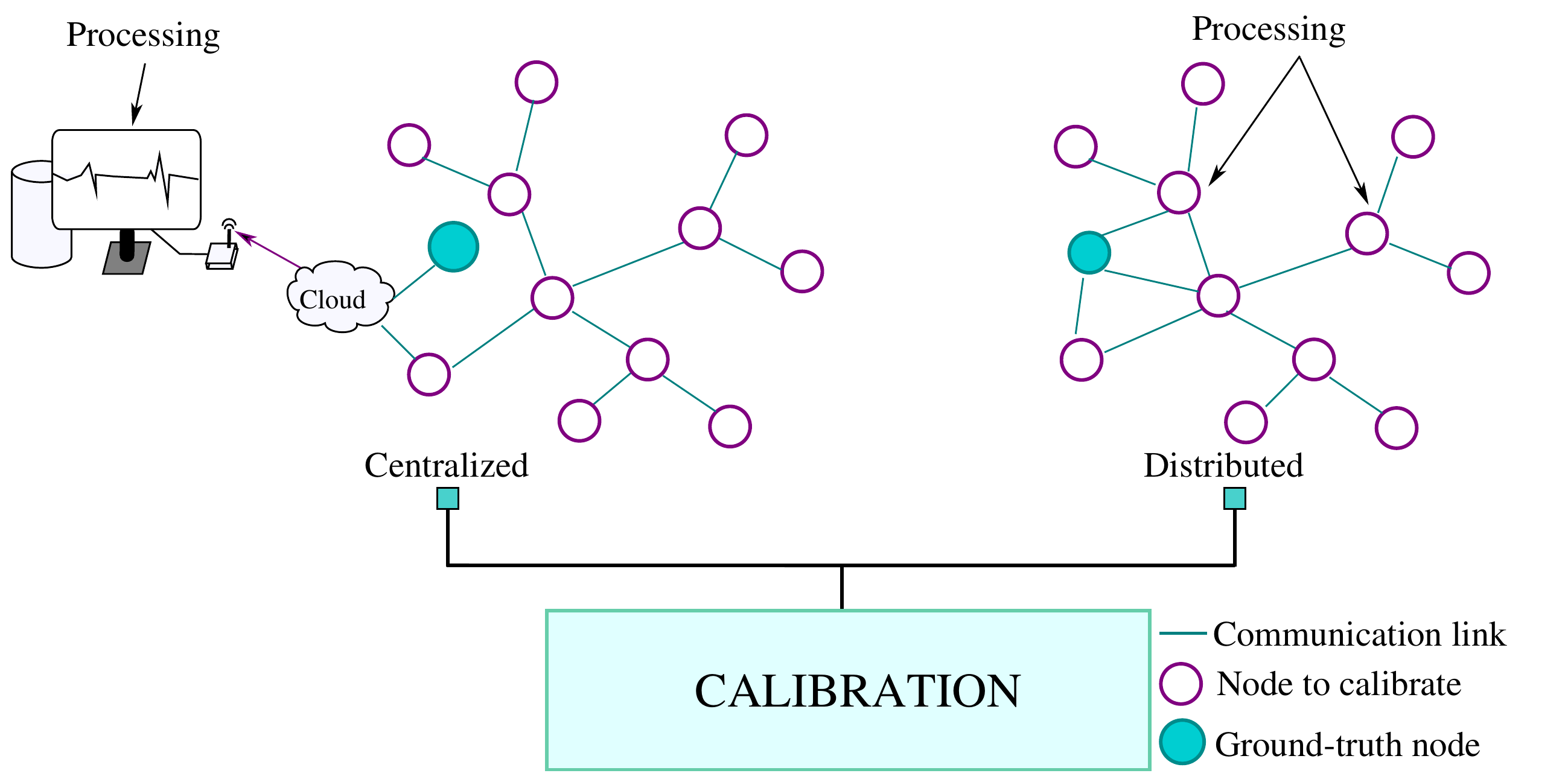}
\caption{Illustration of centralized and distributed calibration in sensor 
networks.}
\label{Fig:Centralized}
\end{figure}
%-%%%%%%%%%%%%%%%%%%%%%%%%%%%%%%%%%%%%%%%%%%%%%%%%%%%%%%%%%%%%%%%%%%%%%%%%%%

%%%%%%%%%%%%%%%%%%%%%%% BEGIN SINGLE TABLE
\begin{table*}[t]
\centering
\caption{Attributes used in calibration approaches classified by applications.
\label{Tab:ATT_single}}
{
\scalebox{0.75}{
\begin{tabular}{|l||cc|cc|ccc|ccc|ccc|cc|cc|l|}
\hline\hline
{\textbf{\rot{References}}}
             & {\textbf{\rot{Micro}}} & {\textbf{\rot{Macro}}}          
             & {\textbf{\rot{Single Sensor}}} & {\textbf{\rot{Sensors Fusion}}}
             & {\textbf{\rot{Non-Blind}}} & {\textbf{\rot{Semi-Blind}}} & {\textbf{\rot{Blind}}}    
             & {\textbf{\rot{Collocated}}} &{\textbf{\rot{Multi-Hop}}} & {\textbf{\rot{Model-based}}}  
             & {\textbf{\rot{Pre/Post}}} &{\textbf{\rot{Periodic}}} & {\textbf{\rot{Opportunistic}}}   
             & {\textbf{\rot{Off-Line}}} & {\textbf{\rot{On-Line}}}  
             & {\textbf{\rot{Centralized}}} & {\textbf{\rot{Distributed}}}  
             & {\textbf{\rot{Application}}}
\\\hline\hline
\multicolumn{19}{|c|}{Attributes used in light, temperature, relative humidity, vibration and accelerometer applications}
\\\hline\hline
%=============================================================================================
%                & Mic & Mac & Sing & Fusion & N-B & S-B & B & Coll & Multi & Sing & Post & Per & Oppr & Off & On & Centr & Distr & App
\cite{Bychkovskiy03} & &  X  &  X   &        &     &     & X &      &   X   &      &   X  &     &      &     &  X &       &  X    & Temperature\\
\cite{Feng03}    &     &  X  &  X   &        &  X  &     &   &      &       &  X   &  X   &     &      &     &  X &  X    &       & Point-lights\\
\cite{Miluzzo08} &  X  &  X  &  X   &        &  X  &  X  &   &      &   X   &      &      &     &  X   &     &  X &       &  X    & Light\\
\cite{Buonadonna05}&X  &     &  X   &        &  X  &     &   &  X   &       &      &  X   &     &      &     &  X &       &  X    & Light, Temp., Hum.\\
\cite{Balzano07} &  X  &     &  X   &        &     &     & X &      &  X    &      &  X   &     &      &  X  &    &  X    &  X    & Temp., Hum.\\
\cite{Wang09}    & X   &     &  X   &        &     &     & X &      &  X    &      &  X   &     &      &     & X  &  X    &  X    & Light\\
\cite{Gao11}     &  X &      &      &  X     &  X  &     &   &      &       &   X  &   X  &     &      &  X  &    &  X    &       & Motion (accelerometer) \\
\cite{Rossini16} & X  &      &  X   &        &     &  X  &   &      &    X  &      &   X  &     &      &     & X  &       &  X    & Light\\
\cite{takruri2010}&    &  X  &  X   &        &     &  X  &   &      &   X   &      &   X  &     &      &     & X  &       &  X    & Temperature \\
\cite{Buadhachain13}& &  X   &  X   &        &     &   X &   &      &       &  X   &   X  &     &      &     & X  &       &  X    & Temperature \\
\cite{Kumar13}   & X  &      &  X   &        &     &  X  &   &      &       &   X  &   X  &     &      &  X  & X  &       &  X    & Temp. Hum., Light \\
\cite{Fujino11,Fujino12,Fujino13}&X&&X&      &     &  X  &   &      &  X    &      &   X  &     &      &     & X  &       &  X    & Temperature\\ 
\cite{Kim2012}   &    &  X   &      &  X     &  X  &     &   &   X  &       &      &  X   &     &      &  X  &    &  X    &       & Vibration (Water Flow)\\
\cite{Lee14}     & X  &      &  X   &        &     &     & X &      &  X    &      &      &     &  X   &     & X  &       &  X    & Temp., Hum., Sound\\
\cite{Wang16}    &    &  X   & X    &        &     &     & X &      &   X   &      &  X   &     &      &     & X  &       &  X    & Temperature
%=============================================================================================
\\\hline\hline
\multicolumn{19}{|c|}{Attributes used in localization, synchronization and target detection applications}
\\\hline\hline
%=============================================================================================
%                & Mic & Mac & Sing & Fusion & N-B & S-B & B & Coll & Multi & Sing & Post & Per & Oppr & Off & On & Centr & Distr & App
\cite{Whitehouse02}&   & X   &      & X      &  X  &     &   &      &       &  X   &  X   &     &      &     & X  &       &  X    & Localization\\
\cite{Tan13}       &   &  X  &      &  X     &     &  X  &   &      &   X   &   X  &  X   &     &      &     & X  &  X    &  X    & Target Detection\\
\cite{Fabeck07}    & X &     &      &   X    &  X  &     &   &   X  &       &      &   X  &     &      &     & X  &       &  X    & Target Detection\\
\cite{Moses2002} & X   &     &      &   X    &     &  X  &   &      &       &  X   &   X  &     &      &  X  &    &  X    &       & Localization\\
\cite{Ihler04,ihler2005}&X&  &  X   &        &     &  X  &   &      &       &  X   &  X   &     &      &     & X  &       &  X    & Localization\\
\cite{Bolognani10} & X &     &  X   &        &     &  X  &   &      &   X   &      &   X  &     &      &     & X  &       &  X    & Localization\\
\cite{Stankovic12a,Stankovic15}
                   &   &  X  &  X   &        &     &     & X &      &   X   &      &      &  X  &      &     & X  &       &  X    & Synchronization\\
\cite{Ando05}      & X &     &  X   &        &     &  X  &   &      &       &  X   &      &  X  &      &     & X  &       &  X    & Synchronization
%=============================================================================================
\\\hline\hline
\multicolumn{19}{|c|}{Attributes used in air pollution and water chemistry applications}
\\\hline\hline
%=============================================================================================
%                & Mic & Mac & Sing & Fusion & N-B & S-B & B & Coll & Multi & Sing & Post & Per & Oppr & Off & On & Centr & Distr & App
\cite{Ramanathan06}  &  X  & &      &   X    &  X  &     &   &   X  &       &      &   X  &     &      &  X  &    &  X    &       & Water Chemistry\\
\cite{hasenfratz2012fly} & X &&     &  X     &  X  &     &   &  X   &   X   &      &      &  X  &  X   &     & X  &       &   X   & O$_3$\\
\cite{spinelle2015,spinelle2017}    
                 & X   &     &      & X      &  X  &     &   &  X   &       &      &  X   &     &      &  X  &    &  X    &       & O$_3$,CO,CO$_2$,NO,NO$_2$\\
\cite{liu2015using}&X&       &      &  X     &  X  &     &   &   X  &       &      &  X   &     &      &  X  &    &  X    &       & CO,CH$_4$,C$_3$H$_8$,CeO$_2$,NiO \\
\cite{Saukh15}   &  X  &     &      &  X     &  X  &     &   &      &   X   &      &      &     &  X   &     &  X &       &   X   & O$_3$,CO\\
\cite{Maag16}    &  X  &     &      & X      &  X  &     &   &  X   &       &      &  X   &     &      &  X  &    &  X    &       &O$_3$,CO,NO$_2$\\
\cite{Esposito2017}&X  &     &      &  X     & X   &     &   &   X  &       &      &  X   &     &      &  X  &    &  X    &       &O$_3$, NO, NO$_2$ \\
\cite{maag2017scan}&X  &     &      &  X     & X   &   X &   &      &   X   &      &  X   &  X  &      &     &  X &       &   X   &O$_3$, CO, NO$_2$\\
\cite{Barcelo2018} &X  &     &      &  X     & X   &     &   &   X  &       &      &  X   &     &      &  X  &    &  X    &       &O$_3$\\
\cite{Tolle05}   &  X  &     &      &  X     &  X  &     &   &  X   &       &      &  X   &     &      &  X  &    &  X    &       & Photosynthetically Active Radiation (PAR)\\
\cite{Xiang15}   &     &  X  &      &  X     &     & X   &   &      &       &   X  &  X   &     &      & X   &    &       &   X   &Air pollution\\
\hline
%=============================================================================================
\end{tabular}
}}
\end{table*}
%%%%%%%%%%%%%%%%%%%%%%%  END SINGLE TABLE

\subsection{Mode of Calibration Processing}
In many situations, the sensor nodes store the sample data or send the measurements 
taken in an interval of time to a centralized node in the sensor network or to a centralized
server on the Internet. After having a set of samples, 
the centralized node or server calculates the calibration parameters (see Figure \ref{Fig:Centralized}). 
In this case, the calibration is called \emph{centralized calibration}. 
The calibration parameters obtained in the centralized server are 
then used to produce calibrated data 
\cite{Buonadonna05,Balzano07,spinelle2017,Esposito2017,Barcelo2018,Kim2012}. 

In \emph{distributed calibration}, a sensor node and its neighboring, or multi-hop,
sensor nodes collaborate to calculate the calibration parameters 
\cite{Miluzzo08,Stankovic15,Wang09,Rossini16,Bolognani10}. 
There are several possibilities for building distributed system architectures. 
The following are some of the approaches.

\subsubsection{Consensus Algorithms}
The consensus problem goal in a distributed system with multiple agents
is the agreement among the agents for a single data value. In the case
of calibration, a network of uncalibrated nodes interacts with calibrated
nodes to solve a consensus problem. The average consensus algorithm measures
the sensor samples' disagreement between the uncalibrated node and a set
of calibrated neighbors. The algorithm converges to a consensus value among
the set of nodes and leads to the discovery of the actual disagreement between
the uncalibrated node's sensor and calibrated nodes' sensors. In a consensus
algorithm,~\cite{olfati2007},
the state of agent $n$ in graph $G$=$(V,E)$ is denoted as $x_n$. A consensus
algorithm to reach an agreement regarding the state of $n$ agents with
dynamics $\dot{x_i}$=$u_i$, where $\dot{x}$ denotes derivative, can be
written as an $nth$-order linear system:
\begin{equation}
\dot{x_i}(t) = \sum_{j\in N_i} (x_j(t) - x_i(t)) + b_i(t),
\end{equation}
with $x_i(0)$=$z_i$$\in$$R$ and $b_i(0)$=$0$. It can be shown~\cite{olfati2007} that the
iterative discrete consensus algorithm can be expressed as
\begin{equation}
x_i(k+1) = (1-\epsilon) x_i(k) + \epsilon\sum_{j\in N_i} \frac{(x_j(k) - x_i(k))}{|N_i|}.
\end{equation}

CaliBree~\cite{Miluzzo08} is 
a model-based calibration network in which ground-truth 
calibrated nodes inform non-calibrated nodes, using a beacon protocol, that they will 
participate in the calibration process. These nodes, then, use a discrete consensus 
distributed algorithm to calibrate the uncalibrated node by computing the degree of
disagreement between the reference nodes and the uncalibrated node. 
The nodes are separated in two sets: nodes that
are calibrated and nodes that are not calibrated.
Let $N_i$ be the neighboring set of
calibrated nodes of an uncalibrated node $i$. 
The state, denoted as $s_i(k)$, is the data sample taken at time $k$, and $\bar{d}_i(k)$ is
the average disagreement measured by node $i$ up to round $k$. The disagreement is defined
as the difference between the state and the consensus value, that is, $d_i(k)$=$s_i(k)$-$\alpha$.
The disagreement at step $k$=$0$, $d_i^{uncal}$, is the difference between the
uncalibrated node $i$ and a calibrated node. The formulation of the disagreement
consensus problem for calibration of data is then~\cite{Miluzzo08}
\begin{equation}
  \bar{d}_i(k+1) =
  \begin{cases}
    (1-\epsilon) \bar{d}_i(k) + \epsilon\sum_{j\in N_i} \frac{(s_j(k) - s_i(k))}{|N_i|} & k>0\\
    d^{uncal}   & k=0
  \end{cases}.
\end{equation}

CaliBree~\cite{Miluzzo08} considered that neighboring calibrated nodes periodically send beacons with
their ground-truth or calibrated sensed data. Uncalibrated nodes run the consensus algorithm to
achieve a consensus value, showing that the algorithm converges to
a minimum disagreement $\bar{d}_i^{min}$. The CaliBree protocol performance is
tested using an uncalibrated mobile light sensor node that comes in the range of calibrated
sensor nodes. 

In a similar work, Bolognani et al. \cite{Bolognani10} proposed a distributed
algorithm to micro-calibrate the radio strength signal indicator (RSSI) that is applied
to localization and tracking for WSNs. The algorithm uses mean squares to
estimate wireless channel parameters, whereas a consensus algorithms is introduced to
estimate the sensor offset that affects the measured received strength of the receiving
node due to fabrication mismatches in the radio chip and that can produce $\pm$6dB in RSSI
measurements.

Stankovic et al. \cite{Stankovic15} proposed a distributed non-blind or blind macro-calibration
mechanism to calibrate the coefficients of a network of $N$ sensors measuring a common
discrete signal $\{x(t)\}$ that is considered a random process. The algorithm
can be considered a non-trivial extension of consensus algorithms. Each sensor $n$
produces an output signal $y_n(t)$ = $\beta_{0n}$ + $\beta_{1n} x(t)$. The goal is to achieve
a calibration function $z_n(t)$ = $a_n y_n$ + $b_n$ = $a_n\beta_{1n} x(t)$ + $a_n\beta_{0n}$ + $b_n$ =
$g_n x(t)$ + $f_n$ so that when there is no ground-truth reference signal
(distributed blind macro-calibration), the sensor produces asymptotically equal
outputs $z_n(t)$. The mechanism may also be adapted to the case in which a sensor
gives ground-truth reference data (distributed non-blind macro-calibration),
producing a gain $g_n$=$1$ and offset $f_n$=$0$ for all $n$=$1,...,N$. If $\theta_n$=[$a_n,b_n$]. 
The distributed blind calibration algorithm minimizes the instantaneous difference between
the signal measured by a sensor and its neighborhood: 
\begin{equation}
J_i = \sum_{j\in N_i} \gamma_{ij} (z_i(t)-z_j(t))^2,
\end{equation}
where $\gamma_{ij}$ are nonnegative scalar weights reflecting the relative importance of
the neighboring nodes. If we apply a {\it stochastic gradient descent algorithm},
the parameter $\theta_n$ is expressed as
\begin{equation}
  \hat{\theta}_n(t+1) = \hat{\theta}_n(t)+\delta(t)\nabla_\theta J_n =
  \hat{\theta}_n(t)+\delta(t) \sum_{j\in N_i} \gamma_{ij} [z_i(t)-z_j(t)]
    \begin{bmatrix}
    y_n(t) \\ 1
  \end{bmatrix},
\end{equation}
where $\delta(t)$ is the step length at each step $t$ of the algorithm. The algorithm
has to begin with an initial point $\theta$(0)=[$1,0$]. At each step, node $n$ has to
receive the updates of its neighboring nodes. Stankovic et al. \cite{Stankovic12a} gave conditions for
the convergence of the algorithm and showed how to modify the algorithm in case a
sensor acts as a ground-truth sensor, giving reference data to its neighbors.

\subsubsection{Gossip Algorithms}
Average consensus gossiping \cite{boyd2006,aysal2009} is an asynchronous protocol where
a node chosen uniformly at random wakes up, contacts a neighbor randomly within its
connectivity radius, and exchanges a state variable to produce a computation update.
The problem is having a network of $n$ nodes that sample an area with initial
values $x(0)$=[$x_1(0),..., x_n(0)$]$^T$ and calculating the average of the entries of
$x(0)$, $\bar{x}=\sum_i x_i(0)$ in a distributed manner. Although average gossip algorithms
perform data fusion (e.g., achieving an average of the samples of the network), gossip algorithms 
allow distributing the computational burden and thus also apply to distributed calibration.

Ramakrishnan et al. \cite{Ramakrishnan11} presented a distributed signature learning and node calibration
(D-SLANC) gossip algorithm for sensor calibration in WSNs. The algorithm relays on distributed
measurements of sensors to estimate source signal's signature and to estimate calibration
parameters using gossip-based distributed consensus. This is a special case of blind
calibration in which there is no a priori knowledge of the signal subspace~\cite{Balzano07}. 
The mechanism can also be classified as micro-calibration, because although there is a
network of sensors that collaborate to estimate the signal, each one of the sensors
will have its own estimated calibration coefficients.
In D-SLANC, $N$ nodes measure a common signal of interest with different gains that are time-invariant.
Each node $n$=$1,...N$, has the following model:
\begin{equation}
  y_n =
  \begin{bmatrix}
    y_{n1}  \\
    \vdots \\
     y_{nK}
  \end{bmatrix}
  =
  \begin{bmatrix}
    \beta^n_{11} & \cdots & \beta^n_{1M} \\
    \vdots       & \ddots & \cdots \\
    \beta^n_{K1} & \cdots & \beta^n_{KM}
  \end{bmatrix}
  \begin{bmatrix}
    x_{1}  \\
    \vdots \\
    x_{M}
  \end{bmatrix}
  +
  \begin{bmatrix}
    \epsilon_{1}  \\
    \vdots \\
    \epsilon_{K}
  \end{bmatrix}
  =
  {\bf A_n}(\beta_n){\bf x}+\epsilon,
\end{equation}
where ${y_i}$$\in$$R^K$ is the noisy observation, ${\epsilon_i}$$\in$$R^m$ is additive
Gaussian noise with zero mean and covariance $\sigma^2\bold{I}$, and $x$$\in$$R^M$ is
the common signal to be estimated.
The whole network can be expressed as follows: 
\begin{equation}\label{eq:SLANC_A}
  \bold{y}=
  \begin{bmatrix}
    {\bf y_1} \\ \vdots \\ {\bf y_N} \\
  \end{bmatrix}
  =
    \begin{bmatrix}
    {\bf A_1}(\beta_1) \\ \vdots \\ {\bf A_N}(\beta_N) \\
  \end{bmatrix}
    \bold{x} +
  \begin{bmatrix}
    \epsilon_{1} \\ \vdots \\ \epsilon_N \\
  \end{bmatrix}
 = \bold{A}(\beta)\bold{x} + \epsilon.
\end{equation}
The signature {\bf $x$} and the calibration parameter {\bf $\beta_n$} are the unknown
parameters to be estimated. Here, there
are no ground-truth data, the sensor measurements act as $y_n$, and the unknowns
are both the calibration coefficients and the true sensing signal (signature estimation).
Since the joint distribution of the unknowns is non-convex, D-SLANC splits the parameter
space into two parts, the signature space and the calibration coefficients space, and
then applies {\it alternating minimization} that consists of performing maximum likelihood
(ML) estimation of one set while keeping the other one fixed. D-SLANC converges to a local
maximum of the likelihood. To perform the signature estimation, a projected-gradient-based
coordinate descent algorithm can be developed in which a distributed consensus
algorithm forms the basis of each step of the coordinate algorithm~\cite{Ramakrishnan11}.
 Using the gradient descent algorithm, each node computes each one of the $M$ components
of the signature signal $x$. At each step of the algorithm it is necessary to calculate
the gradient of the signature. This gradient is obtained using a distributed consensus
mechanism. Ramakrishnan et al. \cite{Ramakrishnan11} provided sufficient conditions to
guarantee convergence of the coordinate descent for the exact consensus and with high
probability for the approximate consensus algorithm. Finally, the local calibration
factors are calculated given the current estimate of the signature.

Buadhachain et al. \cite{Buadhachain13} combined noise compensation using the EM (expectation-maximization) 
algorithm and a gossip-based protocol for sharing calibration information in a network
of temperature sensors. The authors presented a state-space representation for the system 
and a control methodology that calibrates sensor nodes by modifying their output values given 
local sensors' outputs and the models.

\subsubsection{Gaussian Process Regression}
Fujino et al. \cite{Fujino11,Fujino13} proposed to calibrate a network of $N$ thermal sensors
using Gaussian processes. All the sensors measure the same phenomenon, and then
they are neighbors from the point of view of seeing similar correlated data, being
the calibration form of type collocated or single-hop.
The authors considered that a set of sensors acts as ground-truth nodes and proposed
an on-line version of the architecture, instead of the classical off-line. 
The network can be represented by $p(y_i\mid\theta)\sim N(a_i x_i+b_i,\sigma^2)$, 
where $\theta$ consists of all 
sensors' parameters [$a_1$,...,$a_N$, $b_1$,...,$b_N$]. The likelihood can be obtained from $p(y_i\mid\theta)$, 
and the solution of the problem is found using the expectation-maximization (EM) algorithm. 

Gaussian process regression, also known as kriging, is a technique used in geostatistical models 
to predict the value of a function at a given point by computing a weighted average of the known 
values of the function in the neighborhood of the point.
Kumar et al. \cite{Kumar13} used simple kriging to predict the humidity value of a sensor from neighboring
nodes, and the measured drift was then Kalman filtered to get the correct drift estimates. The method
was tested in a wireless network of 54 real sensors, improving RMS error with respect to averaging-based 
prediction models~\cite{krishnamachari2004}. Although ~\cite{Kumar13} used kriging to 
estimate the value of neighboring sensors and then correct the drift, kriging can also
be used as a model-based calibration. Given a set of reference calibrated nodes, the value
of interest can be obtained by interpolation in a given location using kriging. This value
can then be used as the ground-truth value for calibrating a sensor located in the same place.
The accumulated errors now come from the kriging process and from the calibration model used.

\subsection{Combining Calibration Attributes}
The attributes described in section \ref{Sec:Forms} are not disjoint, and it is crucial to define
the system approach in which the calibration will be performed. Multiple approaches can result from
combining the calibration attributes. 
In fact, while describing the calibration attributes, examples of combinations have just appeared.
Some examples are
(i) centralized, collocated, non-blind, micro, off-line, pre-calibration, \cite{Ramanathan06};
(ii) centralized, model-based, non-blind, macro, on-line, pre-calibration, \cite{Feng03};
(iii) centralized, multi-hop, non-blind, micro-calibration, array of sensors, \cite{maag2017scan}; and
(iv) distributed, semi-blind, on-line, micro-calibration, \cite{Rossini16}.
There is no golden rule for selecting a calibration approach for a specific application.
In general, depending on the resources available for the deployment, different combinations
of attributes can yield similar results, being the difference and the quality of information
parameters obtained in terms of calibration errors. As an example, calibrating 
a sensor against a collocated ground-truth sensor node will always produce better 
performance than calibrating the same sensor against a $1$-hop already-calibrated sensor node
or against the value predicted by a kriging process.
Choosing one option or another will depend on the availability of the collocated ground-truth node
or on the availability of other already calibrated nodes.
Table \ref{Tab:ATT_single} describes how the examples in Tables \ref{Tab:LTHR}, \ref{Tab:Local},
and \ref{Tab:AirPoll} combine the attributes, forming complex calibration approaches
for several applications.

%%%%%%%%%%%%%%%%%%%%%%%%%%%%%%%%%%%%%%%%%%%%%%%%%%%%%%%%%%
%      Calibration Guidelines
%%%%%%%%%%%%%%%%%%%%%%%%%%%%%%%%%%%%%%%%%%%%%%%%%%%%%%%%%%
\section{Calibration Guidelines}
\label{Sec:CalGuide} 
In general, there is no defined methodology that indicates how to calibrate the sensors of a network. 
However, in figure \ref{Fig:flowchart}, we show a flowchart that specifies the stages we
recommend to follow in the calibration of a wireless sensor network. These steps are described below:

\begin{figure*}[t]
\centering
\includegraphics[width=.75\textwidth]{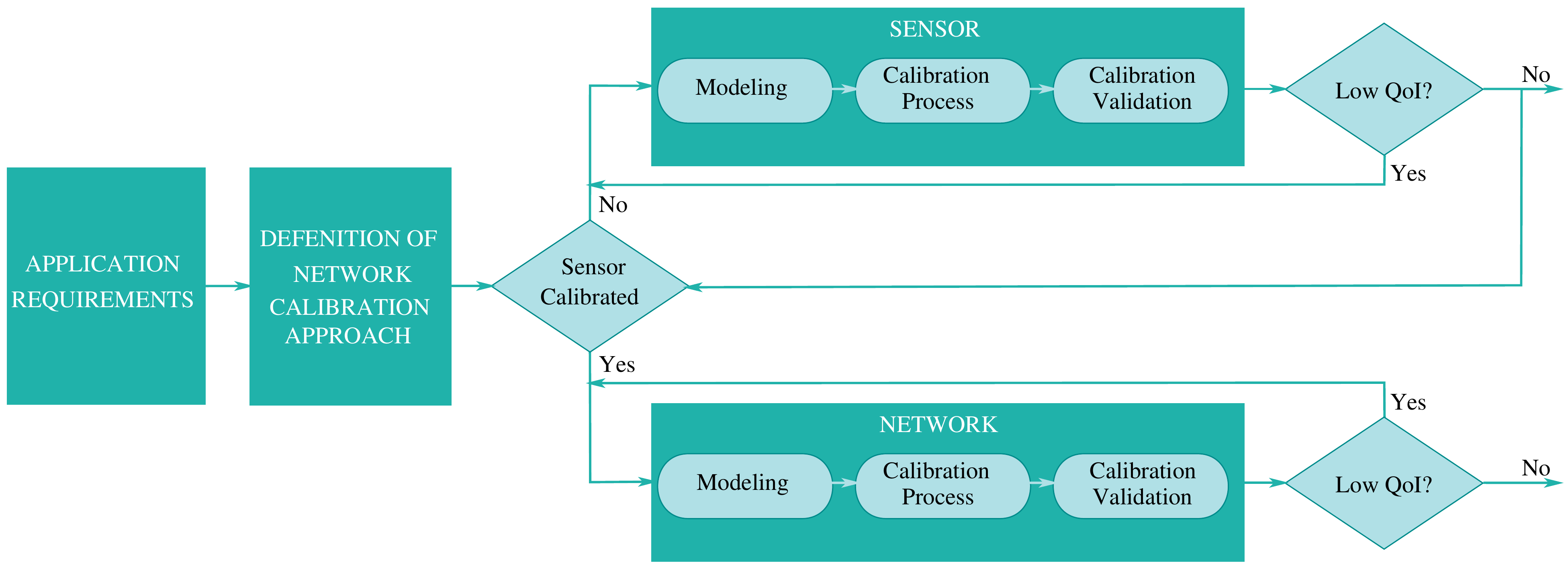}
\caption{Calibration process flowchart.}
\label{Fig:flowchart}
%\vspace*{-15pt}
\end{figure*}

\begin{enumerate}
\item {\bf Application requirements:} This stage defines the intrinsic requirements of the application,
  the ranges of measurement and the QoI metrics. 
\item {\bf Definition of network calibration approach:} In this step, the attributes of the calibration
  are defined. The set of attributes chosen for calibration determines the calibration approach. 
\item {\bf Sensor modeling:} during this stage the mathematical models suitable for the calibration
  of an individual sensor are chosen, such as a linear model or a non-linear model.
  It also analyzes aspects that may influence the calibration such as the time response of the sensor,
  the dynamic range of the sensor or the cross sensitivity.
\item {\bf Sensor calibration process:} at this stage, the size of the training set, the size of the
  test set, or aspects such as the calibration window in case of an opportunistic calibration are defined.
  It is also determined how the environmental conditions or the place of calibration may affect on the place
  where the calibrated data will finally be taken. In addition, it is analyzed if it is necessary to
  recalibrate the sensor due to aspects such as drift or aging. Finally, each sensor is calibrated and its
  QoI is assessed. 
\item {\bf Sensor calibration validation:} in this phase, the individual calibration of the sensors
  is validated. In case of low performance, we go back to the ``sensor modeling'' phase where
  we evaluate if it is necessary to use other models or if there are other aspects that have
  not been considered to influence the calibration.
\item {\bf Network modeling:} during this stage, the appropriate mathematical models are chosen for
  the calibration of the sensor network, such as a consensus or a gossip model.
\item {\bf Network calibration process:} at this point, the parameters participating in
  the network calibration are determined and the information to be exchanged between the nodes
  of the network is specified. Finally, the network is calibrated and its QoI is assessed.
\item {\bf Network calibration validation:} here, the network calibration is validated.
  In case of low performance, we go back to the ``network modeling'' phase and evaluate
 other parameters that may influence the calibration by repeating the process.
\end{enumerate}

%%%%%%%%%%%%%%%%%%%%%%%%%%%%%%%%%%%%%%%%%%%%%%%%%%%%%%%%%%
%      OPEN ISSUES
%%%%%%%%%%%%%%%%%%%%%%%%%%%%%%%%%%%%%%%%%%%%%%%%%%%%%%%%%%
\section{Open Issues}
\label{Sec:Open} 
Although the survey covers two decades of calibration literature in uncontrolled environments, there 
are still issues and challenges that have to be solved. We group the open issues into three categories:
(i) open issues in calibration models deal with aspects related to the calibration models and algorithms
and their effect on sensor calibration, (ii) open issues in calibrating specific sensor technologies
deal with specific aspects of sensor technologies, and (iii) open issues in calibration approaches
deal with aspects related to how combining attributes can produce more complex calibration approaches.

\subsection{Open Issues in Calibration Models}
\label{Sec:Open_CM} 
In the case of calibration using linear and nonlinear models in centralized off-line approaches,
there is not a great knowledge on the size of the data set needed to calibrate the sensors.
In general, the data set is divided into a training and a validation data set. There are few
works that analyze what the minimum size of the data set should be in order to correctly calibrate
a sensor in a real deployment. The minimum size of the data set depends on the kind of sensors and 
sometimes on the environmental conditions. The size of the training set is also important 
in mobile networks in which the sensors are in contact with reference nodes during a window of time
\cite{hasenfratz2012fly}. The time window refers to the number of samples obtained during an interval of time. 
What should be the quality of the calibration in opportunistic or periodic calibration in mobile networks
is another challenge that has to be addressed.

In the past years, there has been an increasing number of papers researching more complex calibration 
techniques based on linear and nonlinear models for centralized, micro, non-blind, collocated, and off-line 
approaches. As previously mentioned, the initial models were based on linear models when the sensors 
were for temperature, humidity, or vibration. Light-point sensors were modeled with splines, and air pollution
with multiple linear regression models. Some sensors present nonlinearities, because their response
function is nonlinear or because of other causes such as drift, aging, or achieving their maximum dynamic
range. Some recent works including \cite{spinelle2015,spinelle2017,Esposito2017,Rossini16} proposed
using deep learning models such as artificial neural networks or support-vector regression.
These models are said to characterize better the nonlinearities than splines but at the cost
of greater computational complexity. Comparative studies on the complexity of these techniques
and models, which ones are appropriate, and how they perform on different families of sensors are needed.

Most calibration techniques have high computational complexity that makes an on-line distributed
approach difficult. High-performance light calibration techniques that can be implemented
on-line in low-computational or in low-capability nodes are needed. Moreover, if the sensors have
nonlinear responses, it is necessary to research how nonlinear models can be applied to on-line
distributed approaches.

Many times there is a lack of ground-truth nodes in the network. One of the possibilities in calibration
in uncontrolled environment networks is to calibrate the sensors using an already calibrated node or 
interpolating (e.g., by kriging or other interpolation schemes) the ground-truth values in a point 
or location and calibrate using these values as reference values. A better knowledge of how the 
error produced by these techniques against a collocated ground-truth is needed.
 
Finally, there is no specific repository dedicated to the calibration of low-cost sensors where one can
find data sets or code used by the calibration research community to allow the comparison of algorithms,
approaches, and calibration methodologies. 

\subsection{Open Issues in Calibrating Specific Sensor Technologies}
\label{Sec:Open_ST} 
There is a lot of research in the communications area on wireless sensor lifetime. However,
this does not take into account the lifetime of the sensor subsystem. Depending on the sensor
application, sensors have different lifetimes. For example, air pollution low-cost sensors for
WSNs based on metal-oxide technology are said to have an average lifetime no longer than
a year and a half, and electro-chemical sensors no longer than a year. Moreover, the drift of these 
sensors is not well known after months of operation \cite{peterson2017}. How these sensors drift,
what their real lifetime is, after how long the sensors should be recalibrated are questions to be researched.
Takruri et al. \cite{takruri2010,takruri2008online} have also investigated the drift in environmental sensors
such as temperature sensors, explaining that detecting drifting sensors and correcting their measurements
would increase the effective life of the network. Wang et al. \cite{Wang16} also stated the difficulty
in separating the detection of drift and calibration process, and mentioned that 
more information such as temporal correlation of sensory data, statistical features, or some other
prior knowledge can be employed to achieve a higher detection rate.

Sensors from the same manufacturer in applications such as air pollution sometimes present
high variability \cite{Barcelo2018,peterson2017}. For example, Peterson et al.
\cite{peterson2017} analyzed metal-oxide sensors and the effect of warm-up, time,
and variability between the same sensors under the same environmental conditions in the
calibration process, showing that sensors have to be calibrated regularly and
highlighting the importance of calibration at a specific location. Learning which sensor
technologies are more stable, what differences exist between sensors from
different manufacturers for the same application, and under what conditions
the sensors have to be calibrated will trigger the deployment of real applications.
Studying temperature and relative humidity sensors, Yamamoto et al. \cite{yamamoto2017} 
reported a mismatch of the calibration error produced in a calibration place and the error
obtained in the prediction in another place with different environmental conditions. 
The same problem has been stated in air pollution
low-cost sensors. Castell et al. \cite{castell2017} mentioned the effect of different
environmental conditions when calibrating NO$_2$ and O$_3$ sensors in a real WSN deployment.
Understanding the different cross-correlations between different phenomena and how
these affect the calibration of a specific sensor still needs to be investigated.

\subsection{Open Issues in Calibration Approaches}
\label{Sec:Open_CA} 
  
Few research studies on multi-hop approaches exist ~\cite{Bychkovskiy03,Saukh15,maag2017scan,Akcan13}.
Multi-hop approaches have the problem of regression dilution. 
Recently, Maag et al. \cite{maag2017scan} have proposed how to minimize or eliminate the regression dilution
effect in multi-hop approaches by using geometric mean regression. Relative to the accumulated error 
in multi-hop approaches, other calibration techniques robust against regression dilution and use of 
a ground-truth node or an already calibrated node as reference in the multi-hop approaches are 
other issues to be investigated.

Most of the works that analyzed blind-calibration used temperature data sets \cite{Balzano07,Lee14,Wang16} 
or synthetic data sets \cite{Stankovic15,Ramanathan06,Lipor14,bilen2014convex}. We believe that although
blind calibration is a promising calibration technique, it has to be tested with wider sets of sensors and
in real scenarios or testbeds. Moreover, knowing the complexity of blind calibration used
in low-computational sensor networks and in distributed approaches is another question to be answered. 

Most of the calibration approaches found in the literature are distributed-based.
However, they are analyzed as centralized-based systems. This is probably because of the small number 
of real testbeds deployed and the difficulty of implementing real-scenario distributed approaches. 
We believe that distributed approaches, 
in general, are an open area that has to be developed and most of the approaches
that are distributed have to be proven that they can work in a distributed manner,
for example, by evaluating their complexity and feasibility in a real testbed.

\section{Conclusions}
\label{Sec:Conclusion} 
A broad range of calibration models was developed to correct measurement 
errors in uncontrolled environments of sensor networks. However, most of the existing calibration 
approaches are built upon a number of assumptions, such as the availability of 
high-quality reference measurements; prior knowledge on the true signal, 
the error model, or both; redundant measurements; spatial correlation, temporal correlation, 
or both; mobility; and interaction or cooperation of nodes. This helps provide some basic 
information required for establishing the calibration process.

Throughout the paper, we have defined the main challenges in the calibration process
in WSNs in uncontrolled environments. These include defining calibration approaches,
calibration models, calibration errors, and the accuracy of the calibration process.
We have first stated that calibration in uncontrolled environments implies calibration
in the field instead of in environmentally controlled chambers with accurate instrumentation.
Nevertheless, this does not mean that accurate instrumentation cannot participate in the calibration process. 
What it means is that the sensor node can interact in the field with nodes that are more accurate or that
were previously calibrated, but sensors are not subjected to the processes of a chamber whose
conditions can be manually controlled.

Moreover, we have described the main models used in calibrating sensors in uncontrolled environments.
These can be divided into linear and nonlinear models. Choosing a specific model
is difficult. In general, this choice depends on the behavior of each sensor. Even sensors of the same family
can behave differently, with some being quite linear and others presenting nonlinearities, 
making the choice of a calibrating model challenging. We have described the typical metrics used to evaluate
QoI in the calibration of sensor networks, emphasizing those most used: $RMSE$, $R^2$, and $MAE$. 

There are a number of attributes that define a specific calibration
approach. Describing these attributes allows us to classify how the calibration
is going to be performed. 
These attributes help identify in which area the calibration is performed, how
many sensors participate in the calibration, whether there is information
available for performing the calibration, what the position of the calibrated
nodes with respect to the uncalibrated node is, at which time the calibration is done, 
whether the network is operative at the time of the calibration, and finally 
if the calibration is done at a specialized node or among all the 
sensor nodes.

Knowing these attributes is the foundation necessary to build complex calibration 
approaches that provide estimation parameters with increasingly sophisticated mathematical
models. Throughout the paper, we have defined and explained how other authors have used these 
attributes to calibrate low-cost sensors in WSNs for different applications including  temperature, 
light-point, vibration, humidity, location, synchronization,
and air pollution (e.g., O$_3$, CO, CO$_2$, NO, and NO$_2$) sensors. 
In summary, we have shown how authors mix calibration attributes to construct
calibration approaches and apply mathematical models to solve 
calibration problems that appear in WSNs.

We think that our paper provides researchers with a comprehensive review of a broad set of
calibration approaches and models applied to WSNs in
uncontrolled environments, while also showing researchers and engineers how to solve
real calibration application problems.\\

%\section{Acknowledgements}
%\label{sec:Adg}
{\bf Acknowledgements}\\
This work is supported by the National Spanish funding TIN2016-78473-C3-1-R, regional 
project 2017-SGR990, and H2020 CAPTOR project.

% Bibliography
%\section{Bibliography}
%\bibliographystyle{elsarticle-num} 
%\bibliography{references}

% that's all folks
\end{document}